\documentclass[twocolumn, tighten, times, twocolappendix]{aastex631}

\begin{document}

\title{The Symbiotic X-ray Binary IGR J16194-2810: \\A Window on the Future Evolution of Wide Neutron Star Binaries From \textit{Gaia}}

\author[0000-0002-1386-0603]{Pranav Nagarajan}
\affiliation{Department of Astronomy, California Institute of Technology, 1200 E. California Blvd., Pasadena, CA 91125, USA}

\author[0000-0002-6871-1752]{Kareem El-Badry}
\affiliation{Department of Astronomy, California Institute of Technology, 1200 E. California Blvd., Pasadena, CA 91125, USA}

\author[0000-0002-6406-1924]{Casey Lam}
\affiliation{Observatories of the Carnegie Institution for Science, 813 Santa Barbara St., Pasadena, CA 91101, USA}

\author[0000-0001-6533-6179]{Henrique Reggiani}
\affiliation{Gemini Observatory/NSF’s NOIRLab, Casilla 603, La Serena, Chile}
\affiliation{Observatories of the Carnegie Institution for Science, 813 Santa Barbara St., Pasadena, CA 91101, USA}



\begin{abstract}

We present optical follow-up of IGR J16194-2810, a hard X-ray source discovered by the {\it INTEGRAL} mission. The optical counterpart is a $\sim500\,L_\odot$ red giant at a distance of $2.1$ kpc. We measured 17 radial velocities (RVs) of the giant over a period of $271$ days. Fitting these RVs with a Keplerian model, we find an orbital period of $P_{\text{orb}} = 192.73 \pm 0.01$ days and a companion mass function $f(M_2) = 0.365 \pm 0.003 \,M_{\odot}$. We detect ellipsoidal variability with the same period in optical light curves from the ASAS-SN survey. Joint fitting of the RVs, light curves, and the broadband SED allows us to robustly constrain the masses of both components. We find a giant mass of $M_\star = 0.99^{+0.02}_{-0.03}\,M_{\odot}$ and a companion mass of $M_{2} = 1.23^{+0.05}_{-0.03}\,M_{\odot}$, implying that the companion is a neutron star (NS). We recover a $4.06$-hour period in the system's TESS light curve, which we tentatively associate with the NS spin period.  The giant does not yet fill its Roche lobe, suggesting that current mass transfer is primarily via winds. MESA evolutionary models predict that the giant will overflow its Roche lobe in $5$--$10$\,Myr, eventually forming a recycled pulsar + white dwarf binary with a $\sim 900$ day period. IGR J16194-2810 provides a window on the future evolution of wide NS + main sequence binaries recently discovered via {\it Gaia} astrometry. As with those systems, the binary's formation history is uncertain. Before the formation of the NS, it likely survived a common envelope episode with a donor-to-accretor mass ratio $\gtrsim 10$ and emerged in a wide orbit. The NS likely formed with a weak kick ($v_{\text{kick}} \lesssim 50$ km s$^{-1}$), as stronger kicks would have disrupted the orbit.

\end{abstract}



\section{Introduction} \label{sec:intro}

 Low-mass X-ray Binaries (LMXBs) are interacting binaries containing a compact object (i.e.\ a neutron star or black hole) accreting from a low-mass donor. ``Symbiotic'' X-ray binaries (SyXBs) are LMXBs in which the donor star is a red giant. While $\sim$200 LMXBs are known in the Milky Way, only a handful of SyXBs have been identified \citep{yungelson_2019, 2023hxga.book..120B}. To date, only two SyXBs have known orbital periods: GX 1+4 ($P_{\text{orb}} = 1161$ d) and 4U 1700+24 ($P_{\text{orb}} = 4391$ d) \citep{hinkle_2006, hinkle_2019, yungelson_2019}. All SyXBs discovered so far are suspected to host wind-accreting neutron stars (NSs) based on the combination of their distinct outburst activity and relatively low average X-ray luminosities ($L_X \sim 10^{32}$ -- $10^{36}$ erg s$^{-1}$) compared to LMXBs in which accretion occurs via Roche-lobe overflow \citep{yungelson_2019}. No SyXBs hosting black holes (BHs) have been discovered to date. We summarize the orbital periods and distances of the six known SyXBs in Table \ref{tab:syxbs}.

The wide orbits of SyXBs present a challenge for binary evolution modeling, because these systems likely survived a common envelope episode --- in which the donor was $\gtrsim 10$ times more massive than the accretor --- without suffering dramatic orbital shrinkage. Furthermore, most NSs are expected to experience a natal kick during supernova, with a typical velocity of $\sim300$ km\,s$^{-1}$ \citep{hobbs_2005}. In most cases, this kick (combined with the mass loss resulting from the supernova) is expected to unbind would-be SyXBs altogether \citep{hills_1983, brandt_1995}. 

Astrometric orbital solutions from the \textit{Gaia} mission \citep[][]{gaia_dr3} have enabled the discovery of three BHs \citep{Gaia_BH1, Chakrabarti_2023, Gaia_BH2, Gaia_BH3} and $21$ NSs \citep{Gaia_NS1, el_badry_2024} orbiting low-mass stars in unexpectedly wide (i.e.\ au-scale) orbits. Since these binaries will eventually evolve into SyXBs \citep{rodriguez_2023}, the challenge of explaining SyXB formation is closely related to the challenge of explaining the characteristics of the population of astrometric compact object binaries discovered by {\it Gaia}.

In this work, we carry out a thorough analysis of the SyXB IGR J16194-2810 in the optical. While the system has been studied extensively at X-ray wavelengths \citep{masetti_2007, ratti_2010, kitamura_2014, bozzo_2024}, the optical counterpart had, until recently, not been studied in detail. Shortly before this paper was completed, \citet{hinkle_2024} reported optical and near-infrared observations of the source. Our analysis here is independent of theirs, but we compare to their work throughout the text. We synthesize archival photometry with new high-resolution optical spectroscopy, light curves, and multi-epoch radial velocities (RVs) to constrain the orbital parameters of the system and derive the stellar properties of the red giant companion. 

The rest of this work is organized as follows. In Section \ref{sec:prior_work}, we discuss the discovery of IGR J16194-2810 and previous work characterizing the system. In Section \ref{sec:source}, we provide an overview of the properties of the X-ray source and its optical counterpart. In Section \ref{sec:data}, we present RV measurements of IGR J16194-2810 collected using the FEROS spectrograph over more than one orbit. In Section \ref{sec:analysis}, we simultaneously fit the ASAS-SN light curves and observed RV curve of the luminous companion with a Keplerian orbit to derive the orbital period and neutron star mass. We also perform spectral energy distribution (SED) modeling of the system's multi-band photometry to study the properties of the giant star. In Section \ref{sec:discussion}, we discuss the implications of our results for the formation of IGR J16194-2810. Finally, in Section \ref{sec:conclusion}, we summarize our main results and consider pathways for future work on SyXBs.

\begin{deluxetable*}{cccc}
\tablecaption{A summary of known symbiotic X-ray binaries, with parameters taken from the literature. In the listed systems, the donor is on the red giant branch or asymptotic giant branch. We do not list systems with supergiant donors (e.g.\ 4U 1954+319), which would more properly be classified as high-mass X-ray binaries (see \citet{hinkle_2020} for a discussion). There are several other unconfirmed associations of red giants with X-ray sources that are not listed here (see \citet{2023hxga.book..120B} for a recent list of SyXB candidates). \label{tab:syxbs}}
\tablehead{\colhead{SyXB} & \colhead{Orbital Period (days)} & \colhead{Distance (kpc)} & \colhead{Reference} \\
\colhead{(1)} & \colhead{(2)} & \colhead{(3)} & \colhead{(4)}}
\startdata
GX 1+4 & $1161 \pm 12$ & 4.3 & \citet{hinkle_2006} \\
IGR J17329-2731 & ? & $2.7^{+3.4}_{-1.2}$ & \citet{bozzo_2018} \\
4U 1700+24 & $4391 \pm 33$ & $0.544 \pm 0.010$ & \citet{hinkle_2019} \\
Sct X-1 & ? & $3.6^{+0.8}_{-0.7}$ & \citet{de_2022} \\
SRGA J181414.6-225604 & ? & $14.6^{+2.9}_{-2.3}$ & \citet{de_mereminskiy_2022} \\
IGR J16194-2810 & $192.73 \pm 0.01$ & $2.129 \pm 0.029$ & This Work \\
\enddata
\end{deluxetable*}

\section{Discovery and Prior Work} \label{sec:prior_work}

IGR J16194-2810 was discovered by \textit{INTEGRAL} and reported in the second IBIS catalog by \citet{bird_2006}. To improve the source localization, \citet{masetti_2007} performed follow-up observations with \textit{Swift}/XRT. They confirmed the association of the X-ray source with a red giant (reporting a chance alignment probability $\sim 4 \times 10^{-5}$), and after obtaining a medium-resolution optical spectrum, characterized the optical counterpart ($V \sim 12.3$) as spectral type M2III. Since the X-ray light curve and spectrum of the source were typical of Galactic LMXBs, they classified IGR J16194-2810 as a new symbiotic X-ray binary hosting a NS. 

Later on, \citet{ratti_2010} used \textit{Chandra} X-ray observations to determine a precise localization of the X-ray source in IGR J16194-2810 to within $0.6$ arcseconds. Further follow-up X-ray observations of the target were also obtained with \textit{Suzaku} by \citet{kitamura_2014} and \textit{Swift}/XRT and \textit{NuSTAR} by \citet{bozzo_2024}. While \citet{kitamura_2014} report the detection of both a soft and a hard emission component from their X-ray spectral analysis, \citet{bozzo_2024} demonstrate that a hot thermal blackbody component with $kT \simeq 1.10$ keV (completed by a power law at high energies) is sufficient to describe the broadband X-ray spectrum. Contrary to the earlier literature, \citet{bozzo_2024} are unable to derive a satisfactory spectral fit involving a Comptonization component, and rule out the presence of a cold thermal component arising from an accretion disk. While they demonstrate that the X-ray spectrum of IGR J16194-2810 is unlikely to undergo significant spectral variability over time, they also find that the source's X-ray brightness varies by a factor of $2$--$3$ over few kilosecond timescales. Furthermore, they point out that the dynamic range of the X-ray luminosity is at least $180$, as the source was significantly brighter in 2007 than it was in 2024.

Recently, \citet{luna_2023} analyzed TESS \citep{TESS} observations of the red giant in IGR J16194-2810. Based on a Lomb-Scargle periodogram analysis \citep{Lomb_1976, Scargle_1982} of the optical data, he reports a detection of a significant periodicity at a period of $242.839$ minutes. He identifies this as the spin period of the neutron star in the SyXB, but finds that the period appears to be transient, since its significance varies both between and within TESS sectors. We consider this measurement suggestive but not definitive (see Section \ref{sec:spin}). 

\citet{kiraga_2012} analyzed ASAS \citep{ASAS} photometry of periodic variable stars coincident with objects from the ROSAT Bright Source Catalog \citep{RASS_BSC}. He detects a variability period of $192.8$ days in the $I$-band light curve of IGR 16194-2810, tentatively attributing it to ellipsoidal modulation. We summarize the physical parameters measured so far in the literature for IGR J16194-2810 in Table \ref{tab:lit_params}.

As we were preparing this manuscript, an independent analysis of IGR J16194-2810 by \citet{hinkle_2024} appeared. The orbital period and most physical parameters we infer in our analysis are in good agreement with their findings. We discuss their work further in Section~\ref{sec:hinkle}.

\begin{deluxetable*}{cccc}
\tablecaption{Physical parameters of IGR J16194-2810 taken from the literature. The provided X-ray luminosities are based on the photogeometric distance and not corrected for absorption. We have applied the zeropoint correction of \citet{lindegren_2021} to the \textit{Gaia} DR3 parallax. \label{tab:lit_params}}
\tablehead{\colhead{Parameter} & \colhead{Description} & \colhead{Value} & \colhead{Reference} \\
\colhead{(1)} & \colhead{(2)} & \colhead{(3)} & \colhead{(4)}}
\startdata
$L_X$ & X-ray Luminosity ($0.3$--$10$ keV) & $0$--$3 \times 10^{34}$ erg s$^{-1}$ & \citet{bozzo_2024} \\
& X-ray Luminosity ($0.2$--$2.3$ keV) & $5.4 \times 10^{32}$ erg s$^{-1}$ & \citet{erosita_dr1} \\
$m_G$ & Apparent G-band Magnitude & $11.421 \pm 0.003$ & \citet{gaia_dr3} \\
$\alpha$ & Right Ascension (X-ray) & $16$h $19$m $33.30 \pm 0.05$s  & \citet{ratti_2010} \\ 
& Right Ascension (Optical) & $16$h $19$m $33.34 \pm 0.03$s & \citet{gaia_dr3} \\
$\delta$ & Declination (X-ray) & $-28^{\circ}$ $07$' $40.3 \pm 0.6$'' & \citet{ratti_2010} \\
& Declination (Optical) & $-28^{\circ}$ $07$' $39.90 \pm 0.02$'' & \citet{gaia_dr3} \\
$\mu_{\alpha} \cos{\delta}$ & Proper Motion in RA & $-0.894 \pm 0.043$ mas yr$^{-1}$ & \citet{gaia_dr3} \\
$\mu_{\delta}$ & Proper Motion in Dec & $-5.287 \pm 0.027$ mas yr$^{-1}$ & \citet{gaia_dr3} \\
$\varpi$ & Parallax & $0.477 \pm 0.036$ mas & \citet{gaia_dr3} \\
$d$ & Distance (Geometric) & $2.101_{-0.151}^{+0.129}$ kpc & \citet{bailer_jones_2021} \\
& Distance (Photogeometric) & $2.131_{-0.116}^{+0.160}$ kpc & \citet{bailer_jones_2021} \\
$P_{\text{spin}}$ & Spin Period & $242.839$ min & \citet{luna_2023} \\
\enddata
\end{deluxetable*}

\section{Source Properties}
\label{sec:source}

\begin{figure*}
\epsscale{1.2}
\plotone{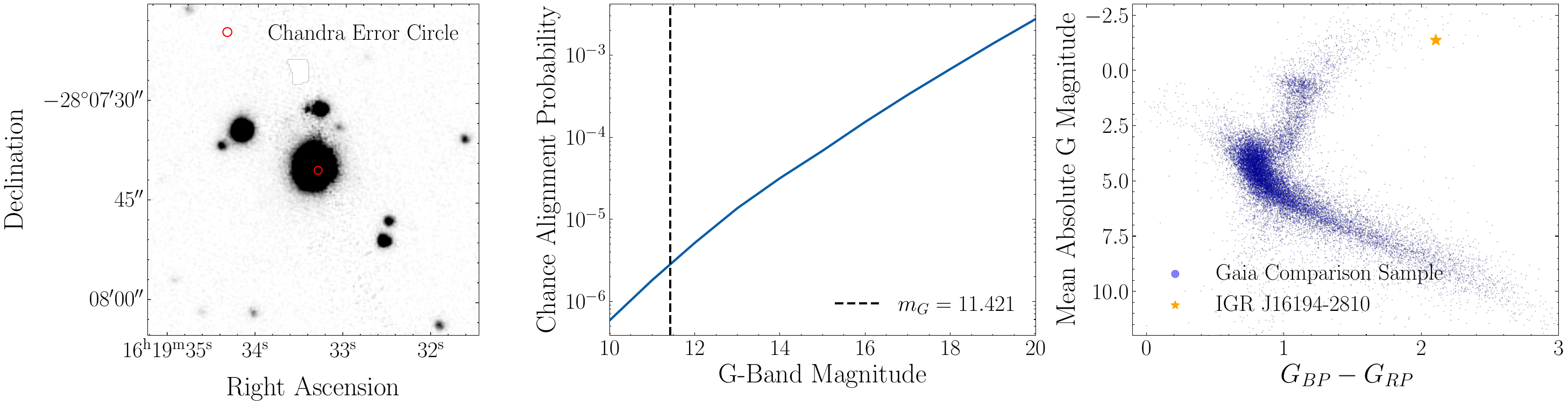}
\caption{Left Panel: PanSTARRS field of view ($50'' \times 50''$) in $g$-band. The image has been rescaled to enhance contrast. The \textit{Chandra} error circle of $0.6$ arcseconds around the X-ray localization is shown. Middle Panel: Chance alignment probability of optical counterparts in the field of IGR J16194-2810 with the X-ray source detected by \textit{Chandra} as a function of $G$-band apparent magnitude. At the apparent magnitude of the giant, the chance coincidence probability of $\sim 3 \times 10^{-6}$ is negligible. Right Panel: Location of IGR J16194-2810 on a extinction-corrected \textit{Gaia} CMD. A sample of stars within $1^{\circ}$ of the target with well-constrained parallaxes is shown for comparison. The companion star in IGR J16194-2810 has evolved far up the red giant branch.}
\label{fig:chance_alignment}
\end{figure*}

We present a PanSTARRS \citep{PanSTARRS} $g$-band image ($50'' \times 50''$) centered on the optical coordinates of IGR J16194-2810 in the left panel of Figure \ref{fig:chance_alignment}. The image has been rescaled to enhance contrast and show the fainter stars surrounding the bright central star. We overlay the $0.6$ arcsecond error circle corresponding to the \textit{Chandra} X-ray localization of \citet{ratti_2010} in red.

To test the probability that the optical counterpart is a chance alignment with the X-ray source detected by \textit{Chandra}, we use the third data release of the \textit{Gaia} mission \citep[DR3;][]{gaia_dr3} to compute the projected number density of stars in a circle of radius $1^{\circ}$ centered on the target. At each apparent magnitude threshold, we multiply the projected number density of stars at least as bright as this threshold by the projected area of the \textit{Chandra} X-ray localization to derive a chance alignment probability. We find that, assuming an X-ray localization error of $0.6''$ based on the \textit{Chandra} pointing uncertainty, the probability that a star with apparent $G$-band magnitude $\leq 11.4$ would be a chance alignment is $\lesssim 3 \times 10^{-6}$. We plot the chance coincidence curve as a function of apparent $G$-band magnitude in the middle panel of Figure \ref{fig:chance_alignment}. 

We display the position of the optical counterpart in IGR 16194-3810 on an extinction-corrected \textit{Gaia} color-magnitude diagram in the right panel of Figure \ref{fig:chance_alignment}. For comparison, we also plot a sample of stars from \textit{Gaia} DR3 that lie within $1$ degree of the target and have well-constrained parallaxes (\texttt{parallax\_over\_error} $> 3$). We obtained extinctions from the 3D dust map of \citet{green_2019}. We filtered out most of the stars that lie below the main sequence using the prescription of \citet{riello_2021}; these stars have anomalous color excess ratios due to contamination from crowding. Clearly, the optical counterpart lies high on the red giant branch.

\section{Data} \label{sec:data}

\subsection{Photometry}

\begin{figure*}
\epsscale{1.2}
\plotone{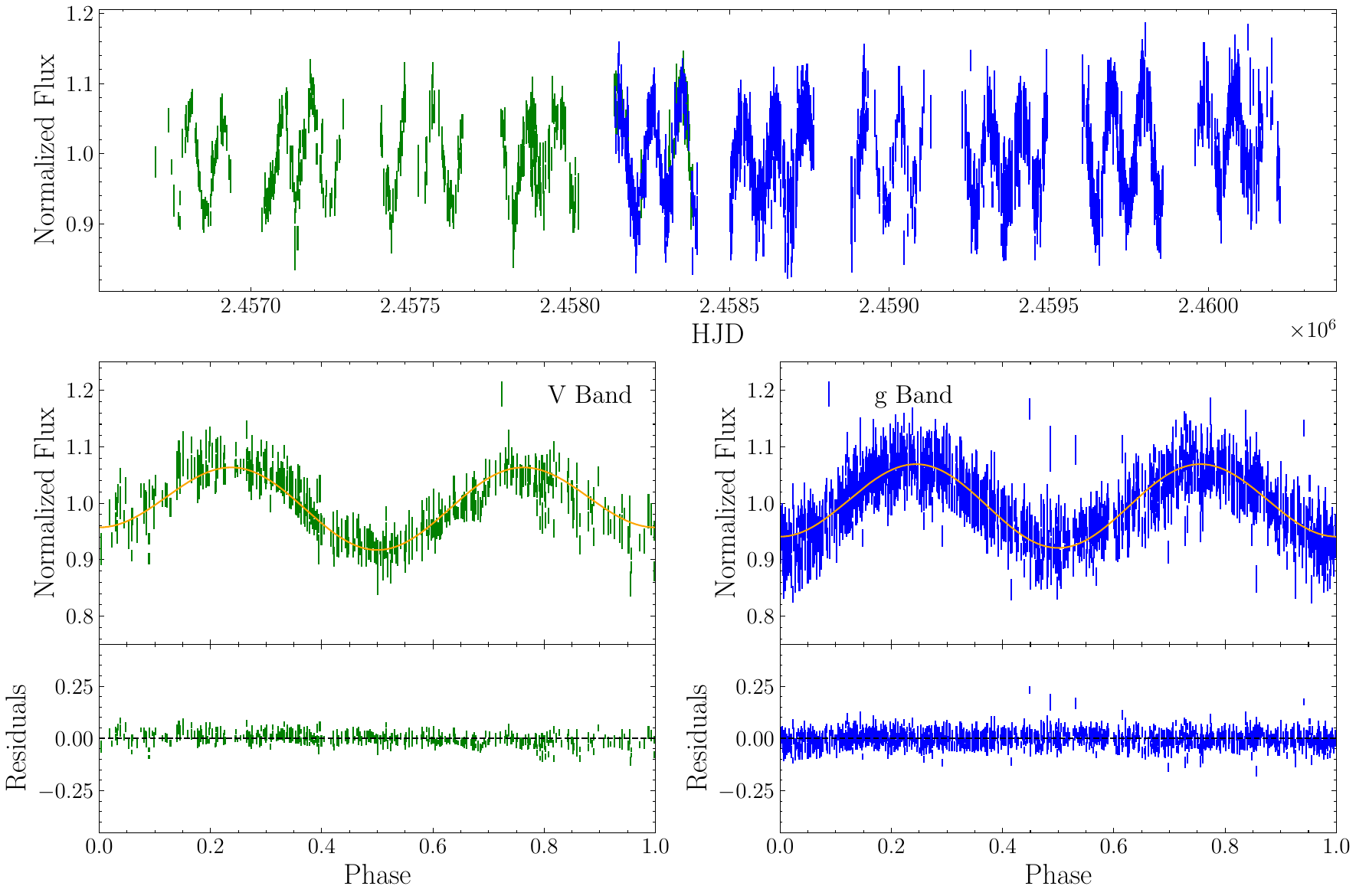}
\caption{In the top panel, we show the $V$-band (green) and $g$-band (blue) light curves of IGR J16194-2810 obtained from the ASAS-SN survey. In the bottom panels, those light curves are phased using the best-fit orbital period of $192.73$ days. The orange curves represent the best-fit ellipsoidal light curve model in each filter. In both filters, the corresponding residuals are consistent with zero, implying that the ellipsoidal light curve model explains almost all of the observed photometric variability.}
\label{fig:lc_fig}
\end{figure*}

We retrieved archival photometry for the red giant in the $V$ and $g$ bands from the ASAS-SN survey \citep{shappee_2014, kochanek_2017}. The unphased ASAS-SN light curves are shown in the top panel of Figure \ref{fig:lc_fig}. From a Lomb-Scargle periodogram analysis, we determine that the ASAS-SN light curves display approximately sinusoidal variability with dominant period $\approx 96$ days. We interpret this as ellipsoidal modulation due to the tidal deformation of the red giant by the neutron star, implying that the true orbital period of the system is $\approx 192$ days. The normalized and phased light curves (with outliers removed) are plotted for both filters in the bottom panels of  Figure \ref{fig:lc_fig}. These light curves are compared to the best-fit ellipsoidal variability models derived from our joint fitting of all available data (see Section \ref{sec:lc_model}).

\begin{figure*}
\epsscale{1.2}
\plotone{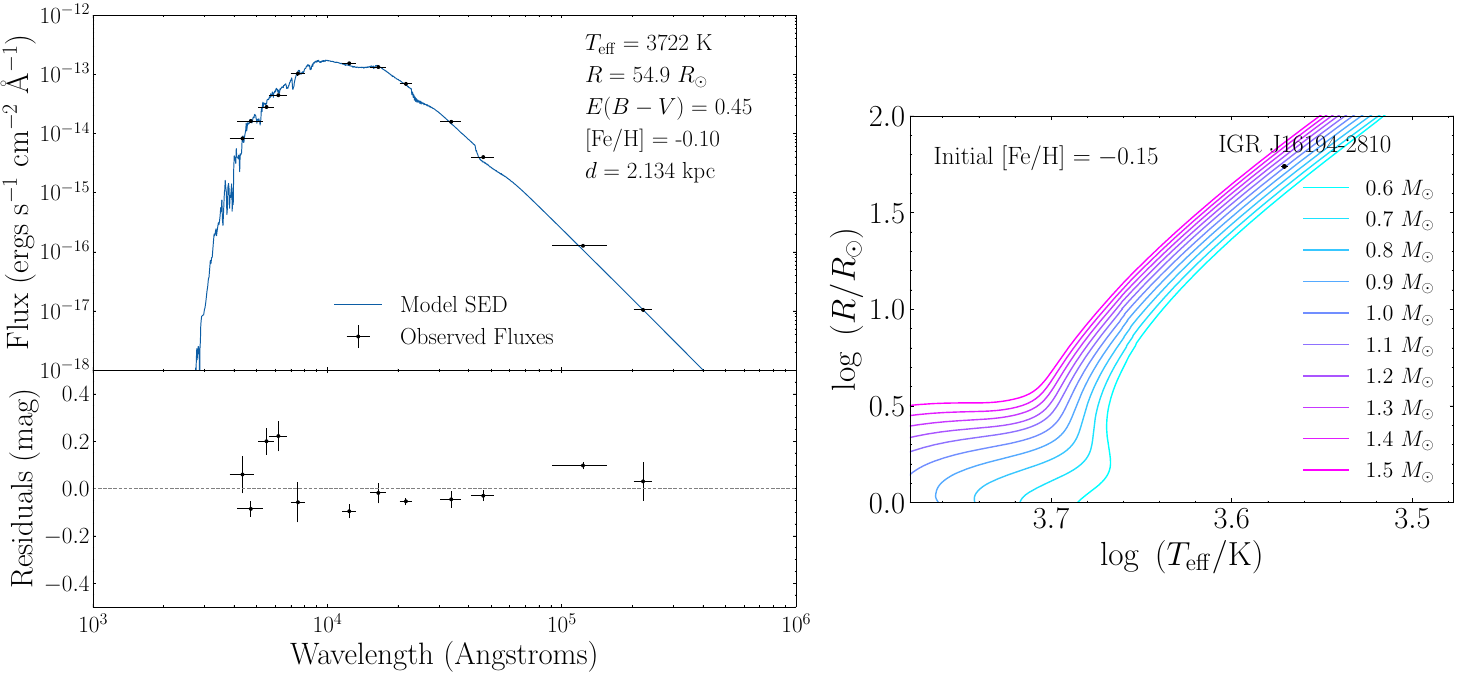}
\caption{Left Panel: Spectral energy density model and associated residuals for the red giant star in IGR J16194-2810. A model SED is plotted based on the MAP parameters of the giant (see Table \ref{tab:derived_params}). The radius and effective temperature inferred from the SED analysis allow us to place tight constraints on the giant's mass. Right Panel: Comparison of the best-fit radius and effective temperature of the red giant to MIST single-star evolutionary models. Assuming an initial metallicity of $\rm [Fe/H] = -0.15$, the location of the red giant on the diagram implies an initial luminous star mass of $\approx 1.0\,M_{\odot}$.}
\label{fig:sed_fig}
\end{figure*}

We retrieved apparent magnitudes in other optical bands from the APASS catalog \citep{henden_2016} and in the infrared from the 2MASS survey \citep{skrutskie_2006} and \textit{WISE} mission \citep{wright_2010}. The source is saturated in deeper surveys such as PanSTARRS \citep{PanSTARRS} and is outside the GALEX footprint \citep{GALEX}. These apparent magnitudes are converted to observed fluxes and are plotted against the central wavelength of their corresponding filters in Figure \ref{fig:sed_fig}. The vertical error bars are derived based on the errors in the apparent magnitudes, while the horizontal error bars represent the effective widths of the corresponding bands. The photometry is compared to the best-fit model SED derived from our joint fitting of all available data (see Section \ref{sec:sed}).

\subsection{Spectroscopy}

\subsubsection{FEROS}
\label{sec:feros}

We observed IGR J16194-2810 17 times with the Fiberfed Extended Range Optical Spectrograph \citep[FEROS;][]{Kaufer1999} on the 2.2m ESO/MPG telescope at La Silla Observatory (programs P111.A-9003, P112.A-6010, P113.26XB). We use exposure times of 900--1200s. The resulting spectra have resolution $R \approx 50,000$ over a spectral range of $350$--$920$ nm. The typical signal-to-noise ratio (SNR) per pixel at 5800\,\AA\, is $\approx 15$.

We reduced the data using the CERES pipeline \citep{Brahm2017}, which performs bias-subtraction, flat fielding, wavelength calibration, and optimal extraction. The pipeline measures and corrects for small shifts in the wavelength solution over the course of a night using simultaneous observations of a ThAr lamp with a second fiber. 

We calculate RVs by cross-correlating a synthetic template spectrum with each order individually and then report the mean RV across 15 orders with wavelengths between 4500 and 6700\,\AA. We calculate the uncertainty on this mean RV from the dispersion between orders; i.e., $\sigma_{\text{RV}}\approx{\rm std}\left({\text{RVs}}\right)/\sqrt{15}$. We used a Kurucz model spectrum \citep{Kurucz1979, Kurucz1993} with $T_{\text{eff}} = 3750$ K, $\log g = 1.5$, and $\rm [Fe/H] = 0$ from the \texttt{BOSZ} library \citep{Bohlin2017} as a template. Adopting a linear limb darkening coefficient based on \citet{claret_bloemen_2011}, we use the direct integration technique implemented by \citet{carvalho_2023} to apply rotational broadening to the template spectrum assuming a projected equatorial rotational velocity of $v \sin i = 15$ km s$^{-1}$ (see Section \ref{sec:vsini}).

All of our measured RVs (with heliocentric corrections applied) are presented in Table \ref{tab:all_rvs}. The median uncertainty of the FEROS RVs is $0.22$ km s$^{-1}$.

\begin{deluxetable}{cc}
\tablecaption{All radial velocities (RVs) measured for IGR 16194-2810 using the FEROS spectra. \label{tab:all_rvs}}
\tablehead{\colhead{HJD UTC} & \colhead{RV (km s$^{-1}$)} \\
\colhead{(1)} & \colhead{(2)}}
\startdata
$2460188.5062$ & $21.49\pm0.15$ \\
$2460224.4984$ & $1.62\pm0.15$ \\
$2460331.3500$ & $0.27\pm0.22$ \\
$2460332.8462$ & $0.51\pm0.17$ \\
$2460333.8497$ & $1.59\pm0.24$ \\
$2460335.8577$ & $2.90\pm0.26$ \\
$2460337.8621$ & $4.51\pm0.25$ \\
$2460339.8638$ & $6.28\pm0.24$ \\
$2460341.8519$ & $7.39\pm0.20$ \\
$2460343.8616$ & $9.09\pm0.26$ \\
$2460370.8610$ & $21.37\pm0.19$ \\
$2460395.7393$ & $17.05\pm0.15$ \\
$2460401.8136$ & $13.05\pm0.17$ \\
$2460424.6068$ & $-4.77\pm0.56$ \\
$2460425.6371$ & $-6.69\pm0.22$ \\
$2460449.6892$ & $-24.42\pm0.32$ \\
$2460459.7927$ & $-28.60\pm0.14$ \\
\enddata
\end{deluxetable}

\subsubsection{MIKE}
\label{sec:mike}

We observed IGR J16194-2810 two times with the Magellan Inamori Kyocera Echelle (MIKE) spectrograph on the Magellan Clay telescope at Las Campanas Observatory \citep{2003SPIE.4841.1694B}. In one observation, we used the 0.7'' slit with an exposure time of 840s, yielding spectral resolution $R \approx 32,500$ over a spectral range of  333--968 nm. In this spectrum, the typical SNR per pixel at 5800\,\AA\, is $\approx 70$. In another observation, we use the 0.35'' slit with an exposure time of 1800s, yielding higher resolution by a factor of $2$, but a lower SNR per pixel at 5800\,\AA\ of $\approx 55$. We reduced the spectra with the MIKE pipeline within CarPy \citep{Kelson_2000, Kelson_2003}. To combine the orders, we perform an inverse variance-weighted average. We normalize the resulting spectrum using a polynomial spline fit to continuum wavelengths.

\section{Analysis} \label{sec:analysis}

\subsection{Light Curve Modeling}
\label{sec:lc_model}

Archival ASAS-SN light curves for the red giant companion in the $V$ and $g$ bands from the ASAS-SN survey are provided in the top panel of Figure \ref{fig:lc_fig}. The normalized and phase-folded light curves, displaying approximately ellipsoidal variability, are provided in the bottom panels of  Figure \ref{fig:lc_fig}. We describe the ellipsoidal variation of the flux of the red giant in a given filter with a two-harmonic model:

\begin{equation}
    F = \bar{F} + \sum_{i = 1}^2 A_{i} \cos{\left(\frac{2 \pi i (t - T_0)}{P}\right)}
\end{equation}

Here, $P$ is the orbital period, $T_0$ is the epoch of conjunction, $\bar{F}$ is the mean flux in the given filter, and $A_i$ is the variability amplitude associated with the $i$th harmonic. We expect $A_2$ to be the dominant component \citep{morris_naftilan_1993, gomel_2021}. We predict $A_2$ for a given choice of mass ratio, Roche lobe filling factor, inclination, and limb- and gravity-darkening coefficients using the prescription of \citet{morris_naftilan_1993} (which assumes circular orbits, tidal locking, and linear limb-darkening and gravity-darkening laws) and apply the correction factor of \citet{gomel_2021} (which is relevant at large Roche lobe filling factors). Unlike \citet{hinkle_2024}, we do not assume that the red giant fills its Roche lobe.

We initially leave $\bar{F}$ and $A_1$ as free parameters, and fit the ASAS-SN light curves separately. We then fix $\bar{F}$ and $A_1$ to the best-fit values derived from this initial fit when performing a joint fit to all available data. We also fix the linear limb-darkening and gravity-darkening coefficients in each filter based on \citet{claret_bloemen_2011}. When predicting $A_2$, the Roche lobe-filling factor is derived from the NS mass, red giant mass, red giant radius, and orbital period. Here, the red giant radius is predicted from the red giant's age, initial mass, and initial metallicity (see Section \ref{sec:joint}), while the NS mass, red giant mass, and orbital period (along with the orbital inclination) are left as free parameters.

\begin{figure*}
    \centering
    \epsscale{1.1}
    \plotone{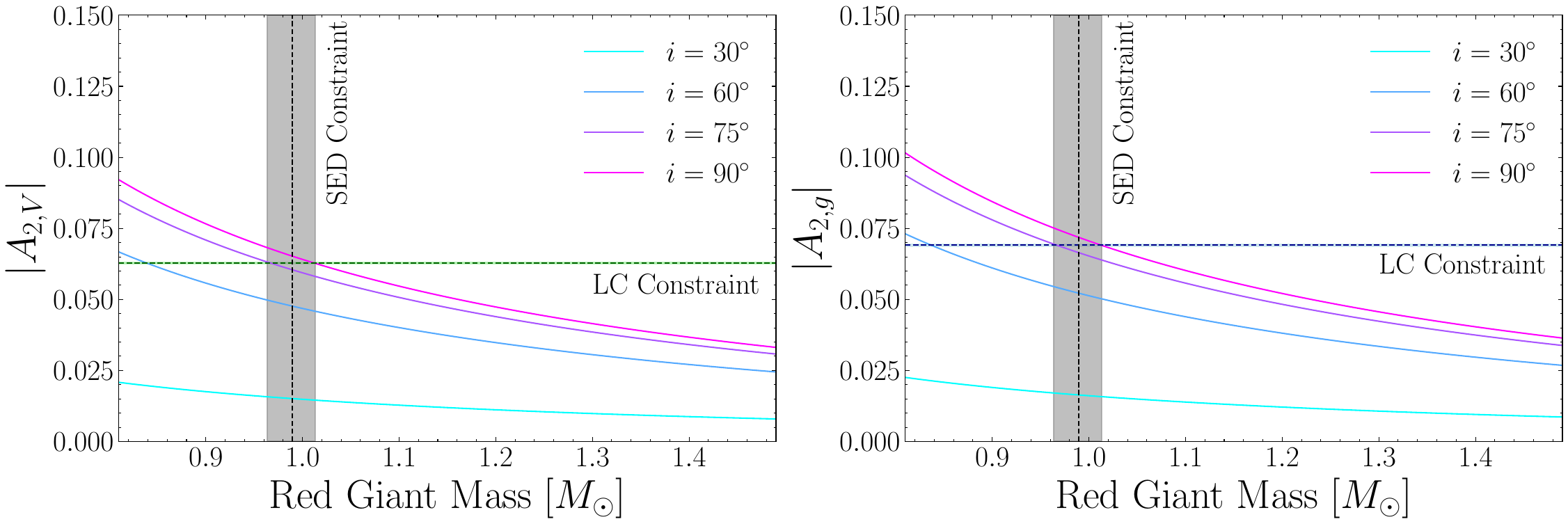}
    \caption{Dependence of the photometric amplitude of ellipsoidal variability (described by $A_2$) on the mass of the red giant and the orbital inclination in both the $V$-band (left panel) and $g$-band (right panel). The red giant radius is fixed to the best-fit value inferred from the SED. As the mass of the red giant increases, its Roche lobe filling factor drops, causing $A_2$ to decrease. In addition, as the orbital inclination decreases, the variability of the geometric cross-section becomes weaker, and $A_2$ tends to decrease. The respective constraints on $A_2$ from the observed light curve and the red giant mass from the observed SED then imply that the orbital inclination must be $>75^{\circ}$; the best-fit value is near $85^{\circ}$.}
    \label{fig:a2_fig}
\end{figure*}

We plot the dependence of the ellipsoidal variability amplitude $A_2$ on the mass of the red giant in both filters (which have different limb-darkening and gravity-darkening coefficients) in Figure \ref{fig:a2_fig}. At larger red giant masses, the primary-to-secondary mass ratio and red giant Roche lobe filling factor both decrease, causing $A_2$ to decrease as well. Furthermore, at more face-on inclinations, the variation in the geometric cross-section of the red giant is not as evident, which also causes $A_2$ to decrease. As a result, the respective constraints on $A_2$ from the observed light curve and the red giant mass from the observed SED imply that the orbital inclination of the system must be greater than $75^{\circ}$. 

We simultaneously fit the ellipsoidal variation in both ASAS-SN filters, assuming the same underlying physical parameters. The log-likelihood term for the light curve data assumes Gaussian uncertainties:

\begin{equation}
    \ln L_{\text{LC}} = - \frac{1}{2} \sum_t \left(\frac{F_{\text{obs}, t} - F_{\text{pred}, t}}{\sigma_{F, t}}\right)^2
\end{equation}

The sum is taken over all observation times and all filters. The best-fit light curve models and their corresponding residuals (taken from the joint fit described in Section \ref{sec:joint}) are plotted in the bottom panels of Figure \ref{fig:lc_fig}. In both filters, the residuals are consistent with zero, implying that ellipsoidal modulation can explain almost all of the observed photometric variability.

\subsection{Fitting the Radial Velocity Curve}

\begin{figure*}
\plotone{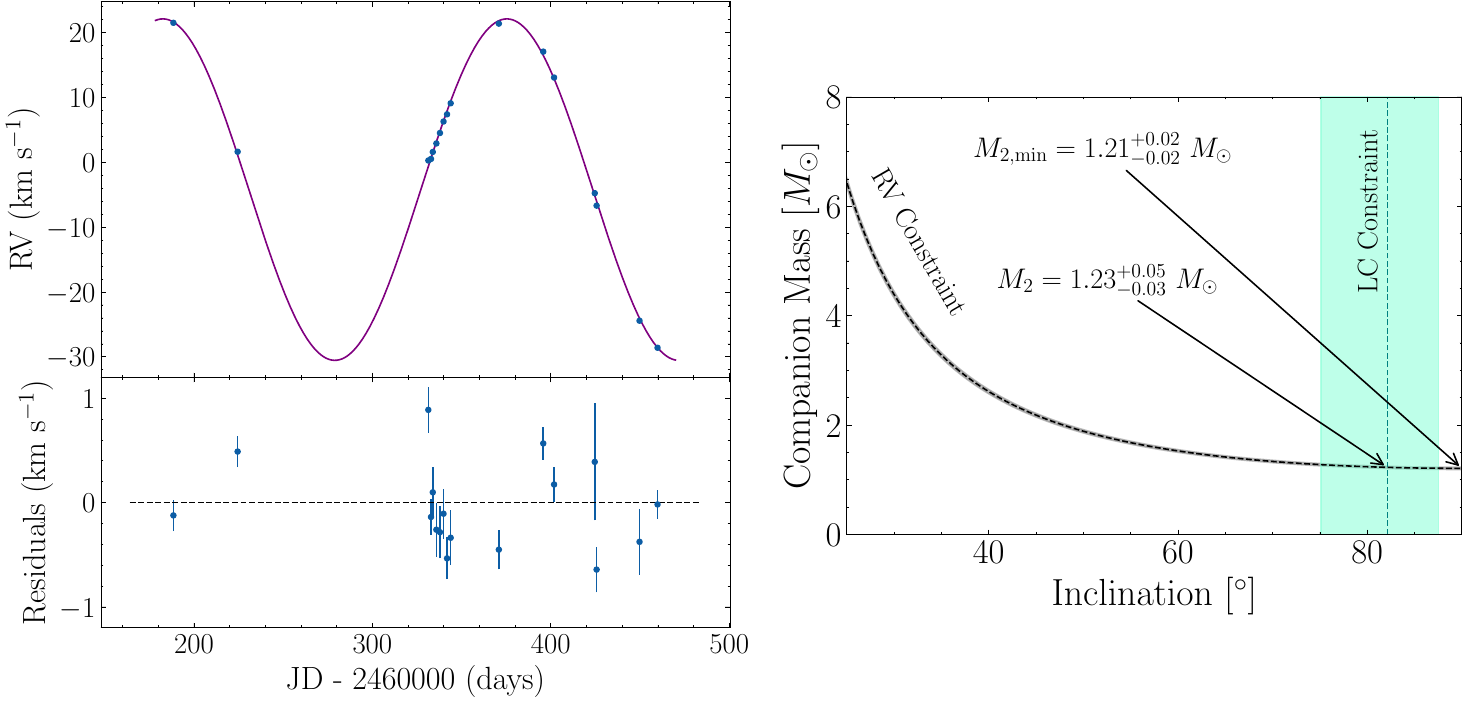}
\caption{Left Panel: Radial velocity (RV) curve and RV residuals of the red giant in IGR J16194-2810. We measured RVs at 17 epochs over more than one orbit using FEROS spectra. The RV semi-amplitude of $26.3 \pm 0.1$ km s$^{-1}$ implies a minimum companion mass of $1.21 \pm 0.02\,M_{\odot}$ (i.e.\ assuming an edge-on orbit), typical of a neutron star. Right Panel: Dynamical constraints on the companion mass. The RVs constrain the companion mass and inclination to the gray shaded region. The inclination constraint from the observed photometric variability amplitude (shaded green region) then implies $M_2 = 1.23_{-0.03}^{+0.05} ~ M_{\odot}$. }
\label{fig:rv_fig}
\end{figure*}

We measured radial velocities (RVs) as described in Section \ref{sec:data}. We plot the resulting RVs in Figure \ref{fig:rv_fig}. We fit the RVs with a Keplerian two-body model:

\begin{equation}
    \text{RV}(t) = \gamma - K \sin{\left(\frac{2 \pi (t - T_0)}{P}\right)}
\end{equation}

Here, $\gamma$ is the center-of-mass RV, $K$ is the RV semi-amplitude, and $P$ and $T_0$ are the orbital period and epoch of conjunction, as before. We assume that the orbit is tidally circularized and synchronized, which is justified for binaries displaying ellipsoidal variability \citep{Zahn_1977, Lurie_2017}. The log-likelihood term for the RV data assumes Gaussian uncertainties:

\begin{equation}
    \ln L_{\text{RV}} = - \frac{1}{2} \sum_t \left(\frac{\text{RV}_{\text{obs}, t} - \text{RV}_{\text{pred}, t}}{\sigma_{\text{RV}, t}}\right)^2
\end{equation}

We present the best-fit RV curve (taken from the joint fit described in Section \ref{sec:joint}), along with its residuals, in the left panel of Figure \ref{fig:rv_fig}. Assuming an edge-on orbit, the observed RV semi-amplitude of $26.3 \pm 0.1$ km s$^{-1}$ implies a minimum companion mass of $1.21 \pm 0.02\,M_{\odot}$. We plot the inferred companion mass as a function of orbital inclination in the right panel of Figure \ref{fig:rv_fig}. If we adopt the inclination constraint implied by the variability amplitudes of the ASAS-SN light curves (see Section \ref{sec:joint}), then we infer a companion mass of $1.23_{-0.03}^{+0.05}\,M_{\odot}$.

\subsection{Spectral Energy Distribution}
\label{sec:sed}

We use \texttt{pystellibs}\footnote{\href{https://mfouesneau.github.io/pystellibs/}{https://mfouesneau.github.io/pystellibs/}} with the BaSeL library \citep{lejeune_1997, lejeune_1998} to generate model SEDs, with the initial mass, initial metallicity, age, and parallax as free parameters. We assume a \citet{cardelli_1989} extinction law with $R_V = 3.1$ and a reddening of $E(B-V) = 0.45 \pm 0.02$ from the \citet{green_2019} 3D dust map. We use \texttt{pyphot}\footnote{\href{https://mfouesneau.github.io/pyphot/}{https://mfouesneau.github.io/pyphot/}} to calculate synthetic photometry. The log-likelihood term for the SED data assumes Gaussian uncertainties:

\begin{equation}
    \ln L_{\text{SED}} = - \frac{1}{2} \sum_{\text{All Filters}} \left(\frac{m_{\text{obs}} - m_{\text{pred}}}{\sigma_{m}}\right)^2
\end{equation}

Our best-fit SED model (taken from the joint fit described in Section \ref{sec:joint}), along with its residuals, is presented in the left panel of Figure \ref{fig:sed_fig}. In the right panel, we use the best-fit radius and effective temperature of the red giant (see Section \ref{sec:joint}) to compare it to MIST single-star evolutionary models \citep{choi_2016}, which imply an initial luminous star mass $M_* \approx 1.0\,M_{\odot}$ at an initial metallicity of $\rm [Fe/H] = -0.15$. As described in Section \ref{sec:joint}, this is the best-fit initial metallicity from our joint fit. The inferred age of the red giant is $9.9 \pm 0.8$ Gyr.

\subsection{Comparison to GALAH Spectra}
\label{sec:galah}

Deriving stellar parameters of cool giants is challenging, as the complex physics of red giant atmospheres involves broad molecular features blended with atomic lines, deviations from hydrostatic equilibrium (i.e.\ giant convective cells), and deviations from local thermal equilibrium \citep{2014dapb.book..217B}. 

To empirically estimate the atmospheric parameters of the red giant companion in IGR J16194-2810 and search for spectroscopic anomalies, we compare a high-resolution MIKE spectrum against the GALAH database \citep{GALAH_2015, GALAH_2021}. Of our two MIKE spectra, we pick the one with lower resolution, but higher SNR. We degrade the MIKE spectrum to $R = 25,000$ and shift all spectra into their respective rest frames. Using the direct integration technique of \citet{carvalho_2023}, we apply a rotational broadening of $v \sin i = 15$ km s$^{-1}$ to each GALAH spectrum (see Section \ref{sec:vsini}). Then, we determine the closest match in pixel space by minimizing the total least-squares deviation between the spectra across all overlapping wavelength ranges that do not include telluric features. Based on the top ten closest matches, we find best-fit stellar parameters of $T_{\text{eff}} = 3710 \pm 80$ K, $\log g = 0.71 \pm 0.14$, $[\text{Fe}/\text{H}] = -0.7 \pm 0.4$, and $V_{\text{mic}} = 2.20 \pm 0.11$ km s$^{-1}$. 

To test the robustness of this approach, we compare our empirical parameters against those derived using the SED-based approach described by \citet{reggiani_2022}. The SED-based effective temperature of $T_{\text{eff}} = 3657 \pm 30$ and surface gravity of $\log g = 1.07 \pm 0.1$ are consistent with our results to within $2\sigma$, while the SED-based metallicity of $\rm [Fe/H] = 0.3 \pm 0.12$ is discrepant. 

We suspect that the GALAH metallicities for luminous giants are systematically underestimated: we find that the mean metallicity of all giants with $T_{\text{eff}}$ and $\log{g}$ within $200$ K and $0.5$ dex of IGR 16194-2810 is $\rm [Fe/H] = -0.5$, which is significantly lower than the expected mean metallicity of $\rm [Fe/H] \sim -0.1$ in the solar neighborhood \citep[e.g.][]{hayden_2015}. As we discuss in Section \ref{sec:joint}, the giant's true metallicity likely falls between the GALAH and SED-inferred values. We note that the \textit{Gaia} XP metallicity inferred by \citet{andrae_2023} is $\rm [M/H] = -0.055$, in line with our expectations. The \textit{Gaia} XP stellar parameters of $T_{\text{eff}} = 3755$ K and $\log{g} = 0.93$ are also in $2\sigma$ agreement with our results.

\begin{figure*}
\plotone{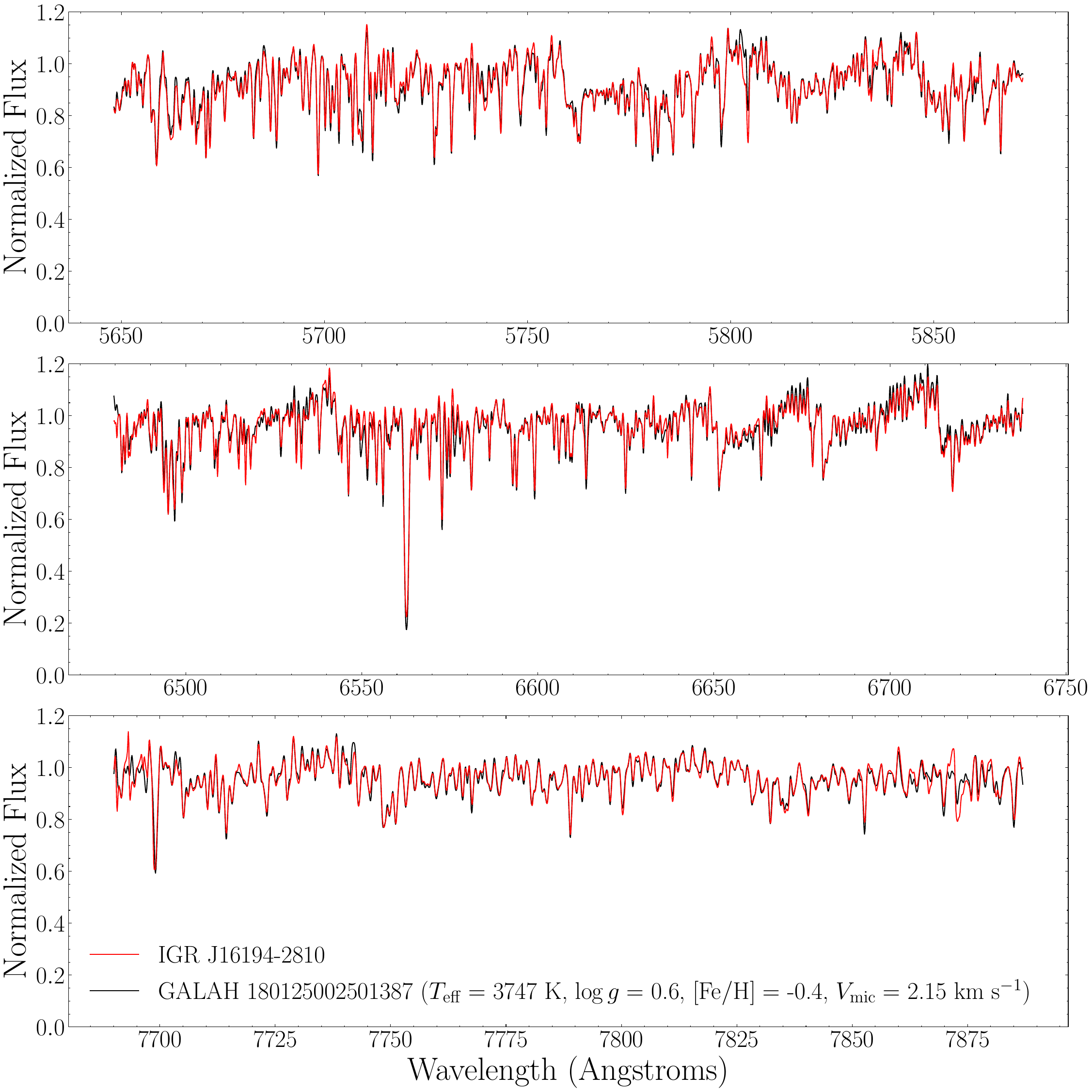}
\caption{Cutouts of a high-resolution optical spectrum of IGR J16194-2810 obtained using the MIKE spectrograph on 02-04-2024. The H$\alpha$ absorption line and molecular band features are evident. In each panel, the MIKE spectrum is compared against the corresponding region of the GALAH spectrum determined to be the closest match. While the two spectra appear to be quite similar, the inferred metallicity is in tension with the value derived for the red giant in IGR J16194-2810 from an SED-based approach.}
\label{fig:mike_galah_fig}
\end{figure*}

We show a high-resolution optical MIKE spectrum of IGR J16194-2810, along with the most similar GALAH spectrum, in Figure \ref{fig:mike_galah_fig}. In the three panels, we compare the normalized and resolution-matched spectra of the two sources in the 2nd, 3rd, and 4th GALAH wavelength windows, respectively. The observed spectrum is typical of a M-type giant, with numerous absorption lines and molecular features evident. Most of the spectrum of IGR 16194-2810 is well-reproduced by its closest spectral doppelgangers. In general, the close match between the two spectra rules out a second luminous star, strong emission lines, or other unusual spectral features. 

We also observe that the red giant's H$\alpha$ absorption line is deeper and blueshifted relative to most similar sources observed by GALAH. We find some sources that have a similar H$\alpha$ absorption line, including the best match displayed in Figure \ref{fig:mike_galah_fig}. This is most likely due to the outward expansion of the chromosphere in these red giants \citep[see e.g.][]{mallik_1993} and unrelated to the companion. 

\subsection{Measuring Projected Equatorial Rotational Velocity}
\label{sec:vsini}

Measuring the projected equatorial rotation ($v \sin i$) of the red giant directly from observed high-resolution spectra provides an independent constraint on the stellar radius and orbital inclination inferred from the joint fitting described in Section \ref{sec:joint}. We begin by selecting a GALAH spectrum with a high similarity score (see Section \ref{sec:galah}) and a low $v \sin i$ as a template for a non-rotating red giant. For comparison, we select the highest-resolution MIKE spectra that we obtained. As before, we degrade the MIKE spectrum to $R = 25,000$ and shift all spectra into their respective rest frames. Then, for each order of the observed spectrum, we use Nelder-Mead optimization \citep{NeldMead65} to find the value of $v \sin i$ that minimizes the least-squares deviation between the spectral order and the rotationally broadened template over the relevant wavelength range. The rotational broadening is applied using the direct integration technique implemented by \citet{carvalho_2023}. Averaging across orders, we derive a projected equatorial rotational velocity of $v \sin i = 17 \pm 2$ km s$^{-1}$ for the red giant, where the error is derived based on the dispersion between orders.

\subsection{Joint Fitting with MCMC Techniques}
\label{sec:joint}

\begin{figure*}
\plotone{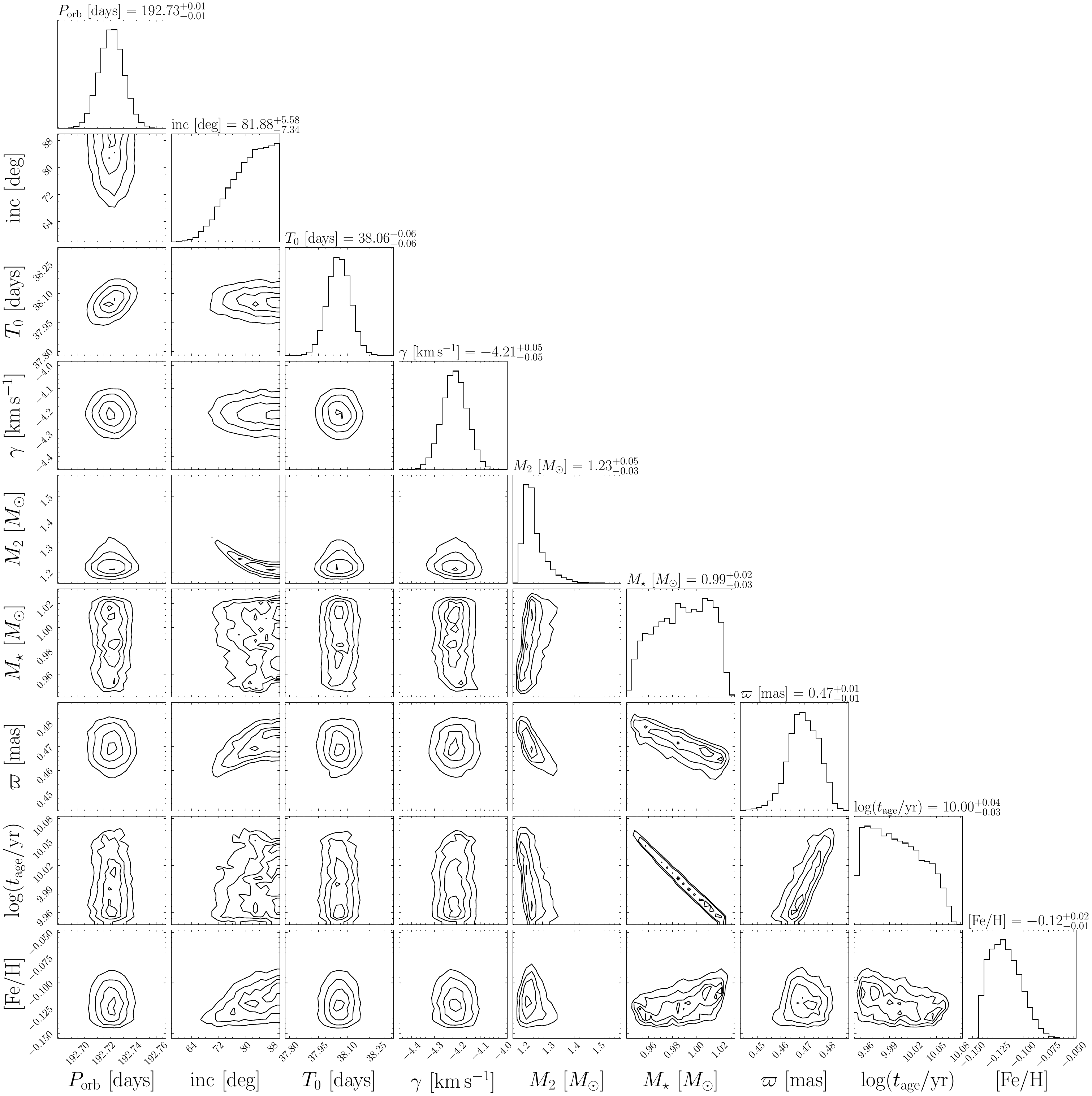}
\caption{Corner plot for orbital and stellar parameters of IGR J16194-2810, assuming an ellipsoidal light curve and two-body Keplerian orbit. The diagonal entries display the marginal distribution of each parameter, while each of the other panels displays a joint distribution. We display the distributions for the present-day mass and surface metallicity (derived using stellar models), as opposed to the initial mass and metallicity (which are directly sampled). The epoch of conjunction $T_0$ is provided relative to JD $= 2460000$.}
\label{fig:corner_plot}
\end{figure*}

\begin{deluxetable*}{ccccc}
\tablecaption{Derived orbital and stellar parameters of IGR J16194-2810. Errors on the median constraints are derived from the 16th and 84th percentiles. Quantities below the dividing line are predicted using the \texttt{isochrones} framework and not directly sampled using MCMC techniques. \label{tab:derived_params}}
\tablehead{\colhead{Parameter} & \colhead{Description} & \colhead{Median Constraint (This Work)} & \colhead{MAP Constraint (This Work)} \\
\colhead{(1)} & \colhead{(2)} & \colhead{(3)} & \colhead{(4)}}
\startdata
$P$ & Orbital Period & $192.73 \pm 0.01$ days & $192.73$ days \\
$i$ & Inclination & $(82^{+6}_{-7})^{\circ}$ & $85.5^{\circ}$\\
$T_0$ & Epoch of Conjunction (JD $- 2460000$) & $38.06 \pm 0.06$ & $38.06$\\
$\gamma$ & Center-of-Mass RV & $-4.21 \pm 0.05$ km s$^{-1}$ & $-4.217$ km s$^{-1}$ \\
$M_{\text{NS}}$ & Neutron Star Mass & $1.23^{+0.05}_{-0.03}\,M_{\odot}$ & $1.223\,M_{\odot}$ \\ 
$\varpi$ & Parallax & $0.470^{+0.007}_{-0.006}$ mas & $0.469$ mas \\
$\log \left(t_{\text{age}}/\text{yr}\right)$ & Red Giant Age & $10.00^{+0.04}_{-0.03}$ & $9.9737$ \\[2pt]
\hline
$M_*$ & Red Giant Mass & $0.99^{+0.02}_{-0.03} \,M_{\odot}$ & $1.0084\,M_{\odot}$ \\ 
$R_*$ & Red Giant Radius & $54.8 \pm 0.7\,R_{\odot}$ & $54.90\,R_{\odot}$ \\
$L_*$ & Red Giant Luminosity & $518^{+13}_{-15}\,L_{\odot}$ & $520\,L_{\odot}$ \\
$T_{\text{eff}}$ & Effective Temperature & $3723^{+6}_{-7}$ K & $3722$ K \\
$\log \left(g/\text{cm s$^{-2}$}\right)$ & Surface Gravity & $0.958^{+0.004}_{-0.009}$ & $0.963$ \\
$\rm[Fe/H]$ & Present-Day Surface Metallicity & $-0.12^{+0.02}_{-0.01}$ & $-0.1026$ \\
$v \sin i $ & Projected Equatorial Rotational Velocity & $14.1 \pm 0.2$ km s$^{-1}$ & $14.37$ km s$^{-1}$ \\
\enddata
\end{deluxetable*}

We simultaneously fit the light curves, RVs, stellar parameters, astrometry, and SED of IGR 16194-2810. The free parameters of our fit are the parallax, orbital period, orbital inclination, epoch of conjunction, center-of-mass RV, and NS mass, along with the red giant's age, initial mass, and initial metallicity. The observed ellipsoidal variability constrains the density and thus the present-day mass of the red giant, with some covariance with orbital inclination. The RVs constrain both the systemic velocity and the companion mass. When taken together, the light curves and RV curve constrain the orbital period and epoch of conjunction. The SED, together with the \textit{Gaia} DR3 parallax, constrains the radius of the red giant. Given the radius and the effective temperature, MIST single-star evolutionary models constrain the red giant's age, initial mass, and initial metallicity. Lastly, the measured projected equatorial rotation constrains the inclination once the radius and orbital period are constrained by other information. 

Unlike \citet{hinkle_2024}, we do not use the lack of eclipses in the optical light curve of IGR 16194-2810 to place a limit on the inclination as a function of mass ratio. This is because --- given the small physical size of the NS and accretion flow relative to a red giant --- we do not expect to see eclipses in SyXBs in the optical bands. Indeed, our joint fit implies $i \gtrsim 75^{\circ}$ and thus predicts an X-ray eclipse. Future targeted observations of IGR J16194-2810 at superior conjunction can test this prediction.

We use truncated uniform distributions to enforce the orbital inclination to lie in $[0^{\circ}, 90^{\circ}]$ and restrict the epoch of conjunction to fall within a time range spanning less than one orbital period. We also place a Gaussian prior on the parallax to IGR J16194-2810 based on the \textit{Gaia} DR3 value after correcting for the parallax zeropoint \citep{lindegren_2021}. 

We use \texttt{isochrones} \citep{2015ascl.soft03010M} to interpolate on MIST stellar models \citep{choi_2016} to predict the red giant's radius, effective temperature, surface gravity, present-day mass, and present-day surface metallicity from its age, initial mass, and initial metallicity. We use a truncated uniform prior to restrict the age to $\leq 13.5$ Gyr. In red giants, the present-day surface metallicity tends to be higher than the star's initial metallicity by $\sim 0.04$ dex due to dredge-up \citep{dotter_2017}.

Assuming Gaussian uncertainties, we add a term $L_{\text{SPEC}}$ to the likelihood to compare the predicted stellar parameters against those derived from the comparison with the GALAH spectra (Section \ref{sec:galah}). Since \texttt{isochrones} predicts a value for the radius of the red giant, this term also includes a comparison of the predicted $v \sin i = 2 \pi R \sin i / P $ with the empirical value from Section \ref{sec:vsini}. In other words:

\begin{equation}
    \ln L_{\text{SPEC}} = \sum_{X \in \{T_{\text{eff}},\, \log g,\, v \sin i\}} - \frac{1}{2} \left(\frac{X - \langle X \rangle}{\sigma_X}\right)^2
\end{equation}

where $\langle X \rangle$ is the best-fit value for parameter $X$ and $\sigma_X$ is the corresponding uncertainty. We do not include the $\rm [Fe/H]$ derived from comparison with the GALAH spectra in this likelihood term due to the reasons discussed in Section \ref{sec:galah}.

We jointly fit all of the available data, using the following total log likelihood:

\begin{equation}
    \ln L = \ln L_{\text{LC}} + \ln{L_{\text{RV}}} + \ln L_{\text{SED}} + \ln L_{\text{SPEC}}
\end{equation}

We use ensemble MCMC sampling \citep[\texttt{emcee};][]{emcee_2013} with $64$ walkers and $5000000$ total iterations to derive best-fit orbital and stellar parameters for IGR J16194-2810. We present the resulting corner plot in Figure \ref{fig:corner_plot}, and report the median and maximum a posteriori (MAP) constraints in Table \ref{tab:derived_params}. We derive a best-fit NS mass of $1.23^{+0.05}_{-0.03}\,M_{\odot}$. Our best-fit orbital period is $192.73 \pm 0.01$ d, in agreement with our Lomb-Scargle analysis. We find an orbit that is close to edge-on, with a red giant of mass $0.99^{+0.02}_{-0.03}\,M_{\odot}$ that is $\approx 82\%$ Roche lobe-filling. The derived effective temperature of $T_{\text{eff}}  = 3723^{+6}_{-7}$ K and surface gravity of $\log \left(g / \text{cm s$^{-2}$}\right) = 0.958^{+0.004}_{-0.009}$ agree with both the GALAH and SED-inferred values to within $\approx 2\sigma$. The best-fit present-day surface metallicity of $\rm [Fe/H]  = -0.12^{+0.02}_{-0.01}$ falls in-between the GALAH and SED-inferred values, as expected. For a red giant age of $9.9 \pm 0.8$ Gyr, the corresponding initial mass and initial metallicity are $1.00^{+0.02}_{-0.03}\,M_{\odot}$ and $-0.16^{+0.02}_{-0.01}$, respectively. Finally, the predicted projected equatorial rotational velocity of $v \sin i = 14.1 \pm 0.2$ km s$^{-1}$ is consistent within $2\sigma$ to the value derived from comparison with the GALAH spectra (see Section \ref{sec:vsini}). We note that the uncertainties on the predicted stellar parameters are implausibly small, and are likely to be dominated by unaccounted-for systematics in the stellar models. Nevertheless, most of our results are in good agreement with those derived in the contemporaneous analysis of \citet{hinkle_2024}.

\subsection{Spin Period Analysis}
\label{sec:spin}

\begin{figure*}
\plotone{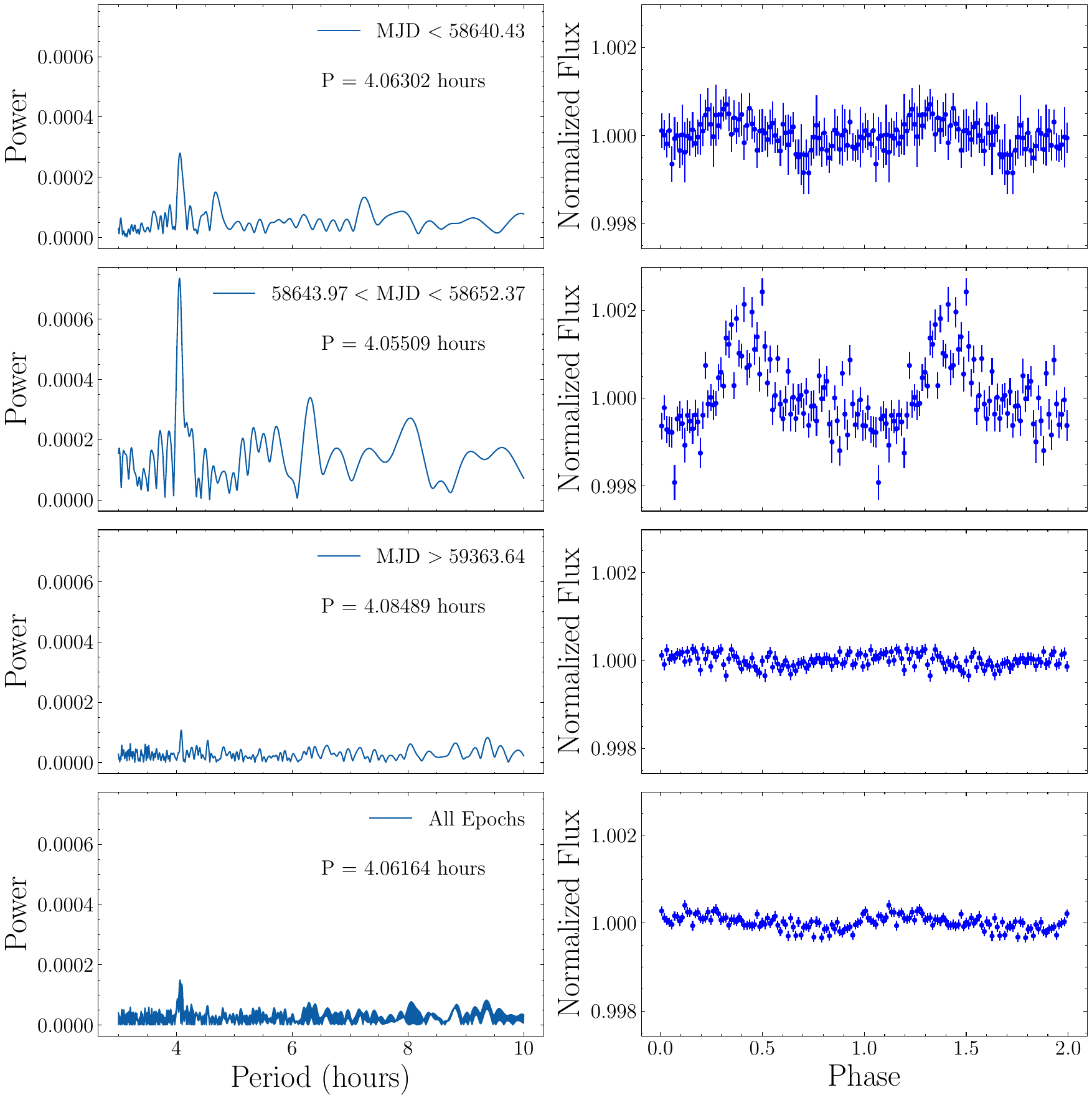}
\caption{Left Column: Lomb-Scargle periodograms for different temporal regions of TESS photometric data. The periodogram is normalized such that the best-fit sinusoid at each frequency has an amplitude of the corresponding power. Right Column: Normalized TESS light curves folded on the corresponding best-fit periods and binned at $80$ bins/cycle. The strongest signal ($P_{\text{spin}} = 4.05509$ hr) is detected in $58643.97 < \text{ MJD } < 58652.37$, where the peak-to-peak variation is about $0.3\%$ of the red giant luminosity.}
\label{fig:tess_lc}
\end{figure*}

\citet{luna_2023} reported a measurement of the NS spin period from TESS data. We attempt to reproduce this measurement and assess how robust it is here. We download high-quality (i.e.\ \texttt{quality\_mask=`hard'}) TESS data for IGR J16194-2810 from Sectors 12 and 39 using the Quick Look Pipeline \citep{tess_qlp}. After removing outliers and de-trending the light curves using a Savitzky-Golay filter, we perform separate Lomb-Scargle periodogram analyses for the entire dataset and over each of the three temporal regions delineated by \citet{luna_2023}. 

We show the normalized TESS light curves for each temporal region and for the entire dataset in the right column of Figure \ref{fig:tess_lc}. In each temporal region, we fold the light curve according to the best-fit period derived from the corresponding Lomb-Scargle periodogram, presented in the left column of Figure \ref{fig:tess_lc}. Each light curve is then binned at $80$ bins/cycle. In the first two regions (i.e.\ the first and second halves of Sector 12), we identify a peak at a period of $\sim 4$ hours. Hints of a weak peak are also detectable in the third region and in the full light curve. While this is suggestive of a NS spin period on this timescale, the weakness and transience of the signal, combined with the lack of a convincing visual pattern in the folded light curve, suggest that the spin period measurement should not be considered to be definitive. Continuous X-ray observations of IGR J16194-2810 over 12+ hours would be useful to confirm the period detection. 

The shape of our folded and binned TESS light curve from the second half of sector 12 does not match that provided by \citet{luna_2023}. The folded light curve shown in his Figure 4 appears significantly less noisy and more coherent than the one yielded by our analysis. We were unable to determine the reason for this discrepancy (Luna, private communication).

The strongest signal is detected in the second half of Sector 12 ($58643.97 < \text{ MJD } < 58652.37$). In that temporal region, we find the best-fit spin period to be $P_{\text{spin}} = 4.06$ hr, with a variation in optical luminosity over one spin period of $\sim 1.5\,L_{\odot}$. The amplitude of this variation is comparable to the X-ray luminosity observed from the X-ray source in IGR J16194-2810 (see Table \ref{tab:lit_params}). While this emission is typical for neutron star accretion flows, it could also arise from an accretion stream or re-processing in the photosphere of the M giant star \citep{jablonski_pereira_1997}.

\subsection{Galactic Orbit}
\label{sec:galactic}

\begin{figure*}
\epsscale{1.2}
\plotone{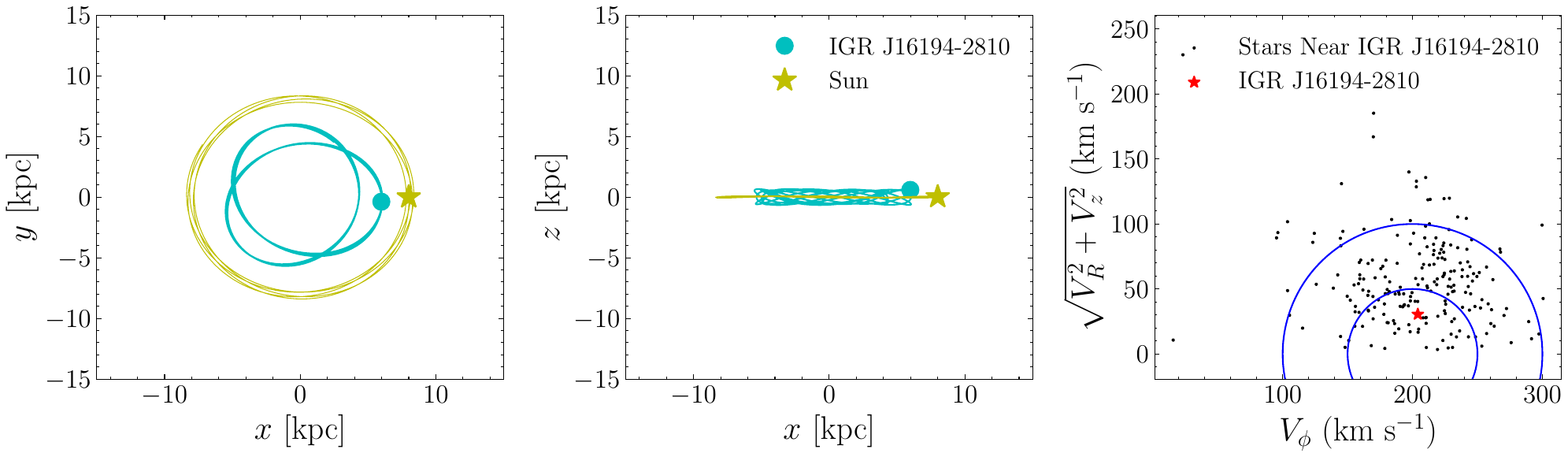}
\caption{Left and Middle Panels: Galactic orbit of IGR J16194-2810 integrated back in time by 1 Gyr based on its current location and spatial velocity. The orbit is clearly puffier and less circular than the solar orbit, shown for reference. This indicates that IGR J16194-2810 was originally orbiting in the plane of the Galactic disk before acquiring a larger spatial velocity due to the supernova of the NS progenitor. Right Panel: Toomre diagram comparing velocities in the direction of the Galactic rotation to velocities perpendicular to the Galactic rotation for stars in the vicinity of IGR J16194-2810. The blue semi-circles are centered on the average rotational speed of $\langle V_{\phi} \rangle = 200$ km s$^{-1}$ and have radii of $50$ km s$^{-1}$ and $100$ km s$^{-1}$. We find that the space velocity of IGR J16194-2810 is consistent with that of nearby stars.}
\label{fig:galactic_orbit}
\end{figure*}

We retrieve the right ascension, declination, and proper motion of IGR J16194-2810 from \textit{Gaia} DR3 (see Table \ref{tab:lit_params}). Combining these initial conditions with our best-fit parallax and center-of-mass RV, and assuming the Milky Way gravitational potential described by \texttt{MilkyWayPotential2022} in \texttt{gala} \citep{gala}, we use $\texttt{galpy}$ \citep{bovy_2015} to integrate the galactic orbit of IGR J16194-2810 back in time by 1 Gyr in the left and middle panels of Figure \ref{fig:galactic_orbit}. 

We find that the galactic orbit of IGR J16194-2810 is puffier and less circular than the solar orbit, but still confined within $\pm 1$ kpc of the disk midplane. This puffiness is possibly due to the natal kick (see Section \ref{sec:formation}).

Under the assumption that IGR J16194-2810 participates in the Galactic rotation, \citet{ratti_2010} derive the peculiar velocity of the system as a function of its center-of-mass velocity (which was unknown at the time). They find an erroneously high minimum peculiar velocity of $280 \pm 66$ km s$^{-1}$. This, in turn, follows from an inaccurate assumption of a high source proper motion of $1.3 \pm 4.7$ mas yr$^{-1}$ in RA and $-20.2 \pm 4.7$ mas yr$^{-1}$ in Dec.\ from UCAC2 --- now superseded by the values provided in \textit{Gaia} DR3 (see Table \ref{tab:lit_params}). Based on their large derived minimum peculiar velocity, \citet{ratti_2010} postulate that the binary is either a halo object or received a natal kick. We have shown that the former is unlikely, and that the latter is not to be expected in the case of a SyXB.

We plot a Toomre diagram of stars in the vicinity of IGR J16194-2810 in the right panel of Figure \ref{fig:galactic_orbit}. We choose stars within a circle of radius 1$^{\circ}$ centered on the target coordinates that have zeropoint-corrected parallaxes within $0.1$ mas of the system and RVs from \textit{Gaia} DR3 with uncertainties $< 5$ km s$^{-1}$. We find that the space velocity of IGR J16194-2810 is consistent with that of nearby stars.

\section{Discussion} \label{sec:discussion}

\subsection{Formation History}
\label{sec:formation}

\begin{figure*}
\plotone{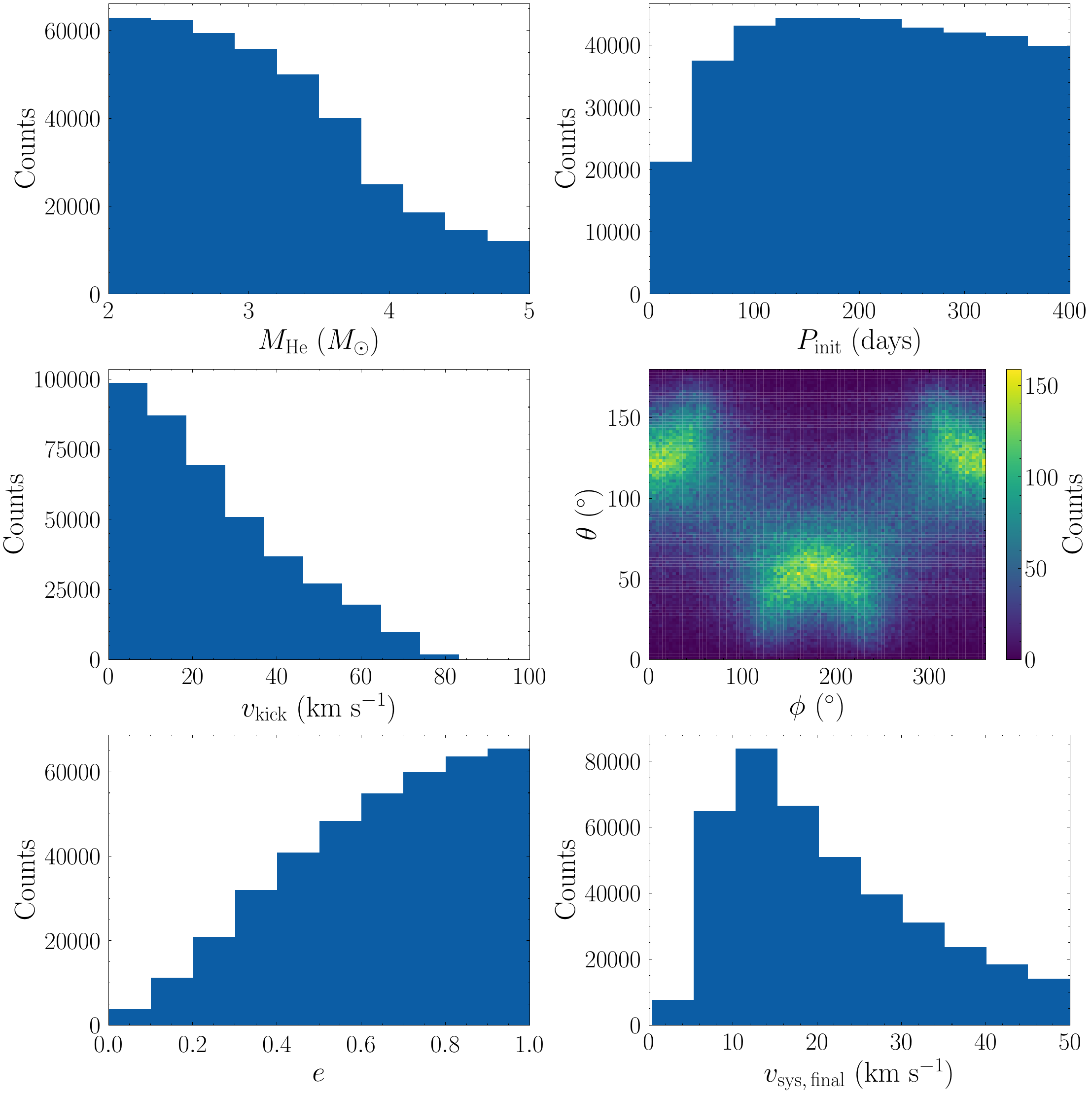}
\caption{Distributions of parameters that allow the final orbit of IGR J16194-2810 to both be bound and as least as wide as observed, assuming $M_{\text{NS}} = 1.4 ~ M_{\odot}$ and $M_2 = 1 ~ M_{\odot}$. Top Row: Distributions of He star masses and initial orbital periods. Middle Row: Distributions of natal kicks and natal kick orientations. Bottom Rows: Distributions of final orbital eccentricities and relative systemic velocities. We find that bound NS + red giant binaries favor small He star masses, large initial orbital periods, and small natal kicks roughly aligned or anti-aligned with the motion of the He star. Prior to tidal circularization, the resulting orbit is expected to be eccentric, with a small relative systemic velocity.}
\label{fig:sn_params}
\end{figure*}

\begin{figure*}
\epsscale{1.1}
\plotone{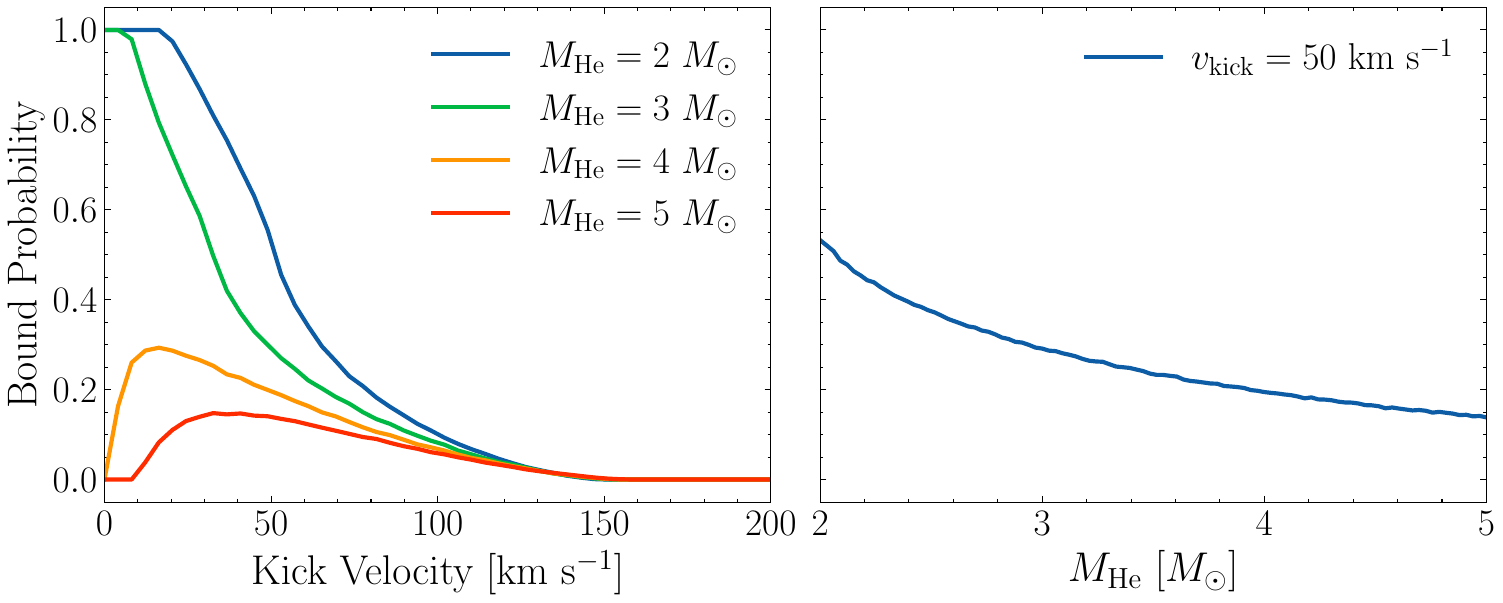}
\caption{Probability that orbit stays bound as a function of natal kick and He star mass. We assume a default initial orbital period of $100$ days, a red giant mass of $1 ~ M_{\odot}$, and a neutron star mass of $1.4 ~ M_{\odot}$. For massive He stars and/or large natal kicks, fine tuning is required to prevent the red giant + NS binary from becoming unbound.}
\label{fig:kick_sim}
\end{figure*}

\begin{figure*}
\plotone{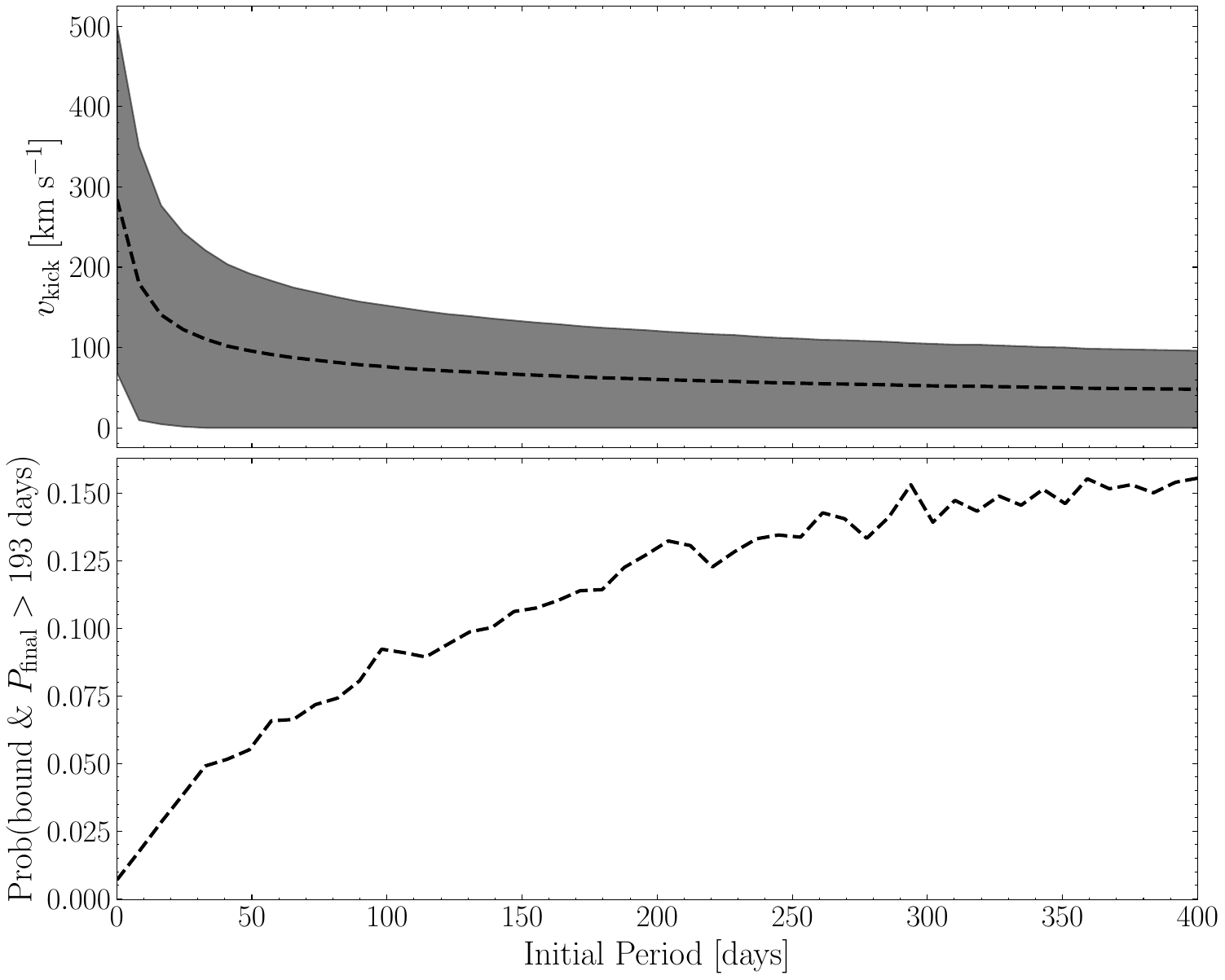}
\caption{Top Panel: Range of natal kicks that allow the final orbit to both be bound and as least as wide as observed as a function of initial orbital period. Bottom Panel: Probability that orbit stays bound and is at least as wide as observed as a function of initial orbital period for median allowed natal kick. We assume a default He star mass of $3 ~ M_{\odot}$, a red giant mass of $1 ~ M_{\odot}$, and a neutron star mass of $1.4 ~ M_{\odot}$. At low initial orbital periods, a large natal kick is required to reach a final orbital separation at least as wide as observed, resulting in a low probability of survival for the resulting binary.}
\label{fig:kick_constraints}
\end{figure*}

As stated in the introduction, SyXBs present a challenge for binary evolution models because they need to survive common-envelope evolution without undergoing a merger event and emerge as a stripped helium (He) star + main-sequence star binary in a relatively wide orbit. Furthermore, the supernova (SN) of the He star will (1) result in significant mass loss and (2) impart a natal kick to the NS that, when taken together, are likely to unbind the binary \citep{hills_1983, brandt_1995, hobbs_2005}. Here, we probe the combination of mass loss, kicks, and initial periods that could reproduce the orbit of IGR J16194-2810. We simulate $10^7$ random combinations of He star masses, initial orbital periods, natal kicks, and kick orientations, and retain the ones that result in a bound orbit at least as wide as observed today. In doing so, we sample from uniform pre-SN distributions of $M_{\text{He}} \in [2 ~ M_{\odot}, 5 ~ M_{\odot}]$, $P_{\text{orb}} \in [0.1 \text{ d}, 400 \text{ d}]$, and $v_{\text{kick}} \in [0 \text{ km s$^{-1}$}, 500 \text{ km s$^{-1}$}]$, respectively. We assume a neutron star mass of $1.4 ~ M_{\odot}$ and a companion mass of $1 ~ M_{\odot}$. Informed by our analysis of the galactic orbit of IGR J16194-2810 in Section \ref{sec:galactic}, we only retain the combinations of initial parameters that result in post-SN relative center-of-mass velocities $\leq 50$ km s$^{-1}$. We do not place a constraint on eccentricity because the orbit was likely circularized by tides when the companion star became a red giant, after the supernova. 

We provide the distributions of the allowed He star masses, initial orbital periods, natal kicks, and kick orientations, along with the resulting eccentricities and relative systemic velocities, in Figure \ref{fig:sn_params}. As expected, we find that a bound orbit wider than $P = 193$ days favors smaller He star masses (i.e.\ smaller amounts of mass loss), larger initial orbital periods, and smaller natal kicks. The $1\sigma$ upper limits on the allowed He star mass and kick velocity are $3.87 ~ M_{\odot}$ and $44.6$ km s$^{-1}$, respectively. We also find that reproducing the observed orbit requires the kick to be either roughly aligned or anti-aligned with the motion of the He star. Peak probabilities occur at $\theta = 120^{\circ}$ or $60^{\circ}$, respectively (due to rotational symmetry). Following the supernova, we expect the resulting orbit to be eccentric with a small relative systemic velocity. However, as mentioned above, tidal interactions will eventually circularize the final orbit.

We plot the probability that the orbit of IGR J16194-2810 remains bound after a supernova as a function of natal kick and He star mass in Figure \ref{fig:kick_sim}. We again follow the prescription of \citet{brandt_1995}, assuming an initial orbital period of $100$ days, a red giant mass of $1 ~ M_{\odot}$, and a neutron star remnant mass of $1.4 ~ M_{\odot}$. At each kick velocity and/or He star mass, we simulate $1000$ random kick orientations and compute the probability that the resulting NS + low-mass star binary remains bound. As expected, the bound probability decreases with increasing He star mass (due to more extensive mass loss) and with increasing kick velocity. 

We simulate the range of natal kicks that allow the final orbit to both be bound and at least as wide as observed as a function of initial orbital period in the top panel of Figure \ref{fig:kick_constraints}. We repeat the analysis in Figure \ref{fig:kick_sim}, except that we assume a fixed He star mass of $3 ~ M_{\odot}$, allow the initial orbital period to vary, and additionally require the final orbit to have $P > 193$ days. We keep track of the minimum and maximum natal kicks that allow for final orbits that satisfy these conditions as a function of initial orbital period. The median allowed kick velocity decreases with increasing initial orbital period, with the allowed range becoming more stringent as well. In the bottom panel of Figure \ref{fig:kick_constraints}, we plot the probability that the final orbit is bound and at least as wide as observed as a function of initial orbital period for the median allowed natal kick. Specifically, we simulate $1000$ random orientations at each median natal kick and report the corresponding probability as a function of initial orbital period. We find that the probability increases as the initial orbital period increases, approaching $\sim 0.16$ as $P_{\text{init}}$ reaches $\sim 400$ days. These simulations suggest that --- given typical NS kick velocities of $\sim300$ km s$^{-1}$ --- a majority of would-be SyXBs become unbound during the supernova, and only systems born with relatively weak kicks survive.

\subsection{Future Binary Evolution}
\label{sec:future}

\begin{figure*}
\epsscale{1.2}
\plotone{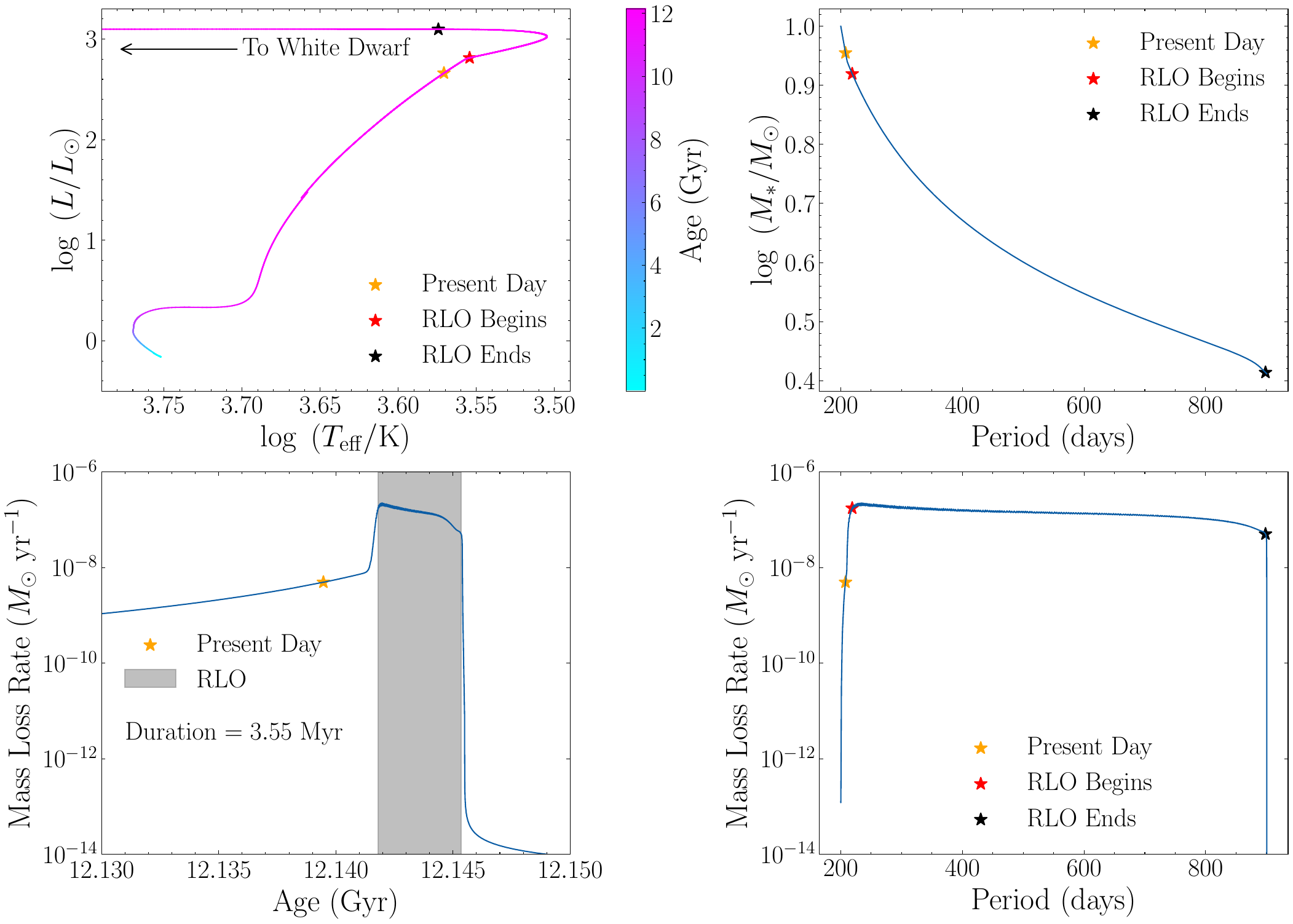}
\caption{Top Left Panel: Evolution of donor star on a H-R diagram, with the luminosity and effective temperature at present day marked with an orange star. We focus on the main sequence and red giant branch. The slope of the red giant branch changes at the onset of Roche lobe overflow (marked with a red star). Top Right Panel: Mass of the red giant star as a function of orbital period. The star loses $\sim 0.6 ~ M_{\odot}$ over the course of its lifetime, the bulk of which is during Roche lobe overflow. This mass loss causes the orbit of the binary to widen significantly, eventually reaching $\sim 900$ days when the red giant loses its envelope and becomes a $0.4 ~ M_{\odot}$ white dwarf. Bottom Left Panel: Mass loss rate as a function of age of the red giant. The Roche lobe overflow phase (which lies about $5.88$ Myr in the future of the binary and is shaded in gray) lasts for around $3.55$ Myr. In the other panels, the start and end of Roche lobe overflow are marked with a red and black star, respectively. Bottom Right Panel: Mass loss rate as a function of orbital period. The elevated mass loss rate during Roche lobe overflow ($\sim 2 \times 10^{-7} ~ M_{\odot}$ yr$^{-1}$) is responsible for the widening of the orbit from $\sim 200$ days to $\sim 900$ days.}
\label{fig:donor_evolution}
\end{figure*}

Based on our derived orbital parameters, we calculate the possible future evolution of the IGR J16194-2810 binary using the Modules for Experiments in Stellar Astrophysics \citep[MESA;][]{paxton_2011, paxton_2013, paxton_2015, paxton_2018, paxton_2019, jermyn_2023}. We start with a $1 ~ M_{\odot}$ star orbiting a $1.4 ~ M_{\odot}$ neutron star (treated as a point mass) in a $200$ day orbit. We use the mass transfer prescription of \citet{kolb_ritter_1990} and the mass transfer efficiency formalism described by \citet{soberman_phinney_1997}. We assume fully non-conservative mass-transfer (i.e.\ $\beta = 1$).

We plot the evolution of the donor star on a Hertzsprung-Russell diagram, focusing on the main sequence and red giant branch, in the top left panel of Figure \ref{fig:donor_evolution}. We find that the red giant in IGR J16194-2810 has not yet achieved Roche lobe overflow (i.e.\ it is currently experiencing mass loss via winds). During the Roche lobe overflow phase, which lasts $\sim 4$ Myr, the red giant experiences a period of heightened mass transfer. This phase is marked by a change in slope of the red giant branch on the H-R diagram. We show the evolution of the mass loss rate as a function of age and orbital period in the bottom left and bottom right panels of Figure \ref{fig:donor_evolution}, respectively. We find that, on average, the red giant experiences a mass loss of $\sim 2 \times 10^{-7} ~ M_{\odot}$ yr$^{-1}$ during Roche lobe overflow. In total, the donor loses around $\sim 0.6 ~ M_{\odot}$ over the course of Roche lobe overflow (top right panel of Figure \ref{fig:donor_evolution}). Eventually, the red giant loses its envelope and descends down the white dwarf (WD) cooling track. The binary ends up with a slowly cooling $\sim 0.4 ~ M_{\odot}$ helium WD orbiting a $1.4 ~ M_{\odot}$ NS in a $\sim 900$ day orbit, comparable to several of the longest-period WD + millisecond pulsar binaries known \citep[e.g.][]{tauris_2011}. 

An upper limit for the accretion rate of the NS today (assuming $v_{\text{wind}} \gg c_s$) is given by the Bondi-Hoyle-Lyttleton (BHL) equation:

\begin{equation}
    \dot{M}_{\text{BHL}} = \frac{G^2 M_{\text{NS}}^2 \dot{M}_{\text{wind}}}{v_{\text{wind}}^4 a^2}
\end{equation}

\noindent where $a$ is the separation between the NS and its red giant companion. Approximating $v_{\text{wind}}$ as the photospheric escape velocity of the red giant (about $84$ km s$^{-1}$ assuming our best-fit parameters) and adopting the present-day mass loss rate from our MESA simulation (about $4.94 \times 10^{-9}\,M_{\odot}$ yr$^{-1}$) for $\dot{M}_{\text{wind}}$, we find that the predicted average accretion rate today is approximately $1.6 \times 10^{-10} ~ M_{\odot}$ yr$^{-1}$. Assuming an accretion efficiency of $\eta = 0.1$, the corresponding accretion luminosity (produced mainly in X-rays) is $L_X = \eta \dot{M}_{\text{BHL}} c^2 \approx 10^{36}$ erg s$^{-1}$. At a distance of about $2.1$ kpc, this implies an observed X-ray flux of $F_X \approx 2 \times 10^{-9}$ erg s$^{-1}$ cm$^{-2}$. Our predicted accretion luminosity is about two orders of magnitude higher than the values reported in Table \ref{tab:lit_params}. This could be due to a reduced accretion rate, reduced accretion efficiency, or both. Indeed, \citet{yungelson_2019} point out that, for slow NS rotation, the NS's magnetic field causes the gravitationally captured plasma to form a hot, convective shell above the magnetosphere. This plasma enters the magnetosphere at a steady rate regulated by plasma cooling (due to the Rayleigh-Taylor instability), sub-sonically settling down with reduced accretion rate $\dot{M} \lesssim 0.5\,\dot{M}_{\text{BHL}}$.

\subsection{Comparison of SyXBs to the Population of Wide NS + MS Binaries}

Astrometric orbital solutions from \textit{Gaia} DR3 have enabled the discovery of $21$ NS candidates orbiting low-mass stars in wide orbits \citep{Gaia_NS1, el_badry_2024}. Based on our derived results, we can compare the space density and implied birth rate of SyXBs to that of the population of wide NS + MS binaries. At a distance of about $2.1$ kpc, IGR J16194-2810 is probably the nearest SyXB with $P_{\text{orb}} \lesssim 1000$ days (i.e.\ detectable in \textit{Gaia} DR3). This suggests a space density $\sim 100$ times lower than what is estimated for the \textit{Gaia} NS + MS binaries, the closest of which is $\sim250$ pc away from the Sun \citep{el_badry_2024}. However, the X-ray bright lifetime of SyXBs is likely only on the order of $\sim 10$ Myr (see Section \ref{sec:future}). Interpolating on MIST stellar models with \texttt{isochrones}, we compute the middle 68\% of the main sequence lifetimes of the companions of the NS candidates reported by \citet{el_badry_2024} to be approximately $6$--$9$ Gyr: almost 1000 times longer than the X-ray bright lifetime of SyXBs. 

Taken together, these space densities and lifetimes suggest a comparable or higher birth rate for SyXBs relative to the \textit{Gaia} NS + MS binary population. Given that the \textit{Gaia} sample is not yet complete, it appears likely that most SyXBs are descendants of the wide NS + MS binaries being revealed by \textit{Gaia}. 

A caveat to this analysis is that we are probably still missing some NS + MS binaries. The upcoming \textit{Gaia} DR4 is expected to lead to more discoveries and shed further light on this population \citep{nineties_trend}. In addition, the SyXB 4U 1700+24, which has an orbital period of 4391 days \citep{hinkle_2019}, lies at a distance of just $\sim0.4$ kpc \citep{masetti_2002}. This suggests that NS + MS binaries with orbital periods too long to be detectable by \textit{Gaia} may be even more common then the ones we consider here.

\section{Conclusion} \label{sec:conclusion}

IGR J16194-2810 is a confirmed member of the rare class of symbiotic X-ray binaries, in which a compact object accretes from an evolved low-mass star. IGR J16194-2810 has been studied extensively in the X-rays and was suspected to host a neutron star based on its X-ray phenomenology. In this work, we present a dynamical mass measurement of the neutron star with optical spectroscopic follow-up. We synthesize our observations with archival photometry to determine the orbital period of the system, which is 192.73 days. All our inferred physical properties of the system are listed in Table~\ref{tab:derived_params}. We summarize our main results below.

\begin{itemize}
    \item The optical counterpart to IGR J16194-2810 is high on the red giant branch  (Figure \ref{fig:chance_alignment}). Given the giant's apparent magnitude of $G = 11.4$, the chance alignment of the optical counterpart in the field of IGR J16194-2810 with the X-ray source detected by Chandra is $\lesssim 3 \times 10^{-6}$. This, combined with the fact that we find the companion to be a dark object with mass $\approx 1.2\,M_{\odot}$ and the giant to be nearly Roche lobe-filling, confirms beyond reasonable doubt that the system is a SyXB. 
    
    \item Based on ASAS-SN optical light curves displaying ellipsoidal modulation on a $\approx 96$ day variability period (Figure \ref{fig:lc_fig}), we infer that the red giant fills $\approx 82\%$ of its Roche lobe. We confirm a tentative detection of a spin period of $4.055$ hours in the second half of Sector 12 TESS data (Figure \ref{fig:tess_lc}), but do not consider this measurement to be definitive.
    
    \item By fitting an SED model to archival photometry of the red giant companion (Figure \ref{fig:sed_fig}), we derive its stellar parameters to be $T_{\text{eff}} = 3723^{+6}_{-7}$ K, $\log \left(g / \text{cm s$^{-2}$}\right) = 0.958^{+0.004}_{-0.009}$, and $\rm [Fe/H] = -0.12^{+0.02}_{-0.01}$ (Table \ref{tab:derived_params}). These values agree with those derived from our comparison of the high-resolution optical MIKE spectrum to the GALAH spectral database to within 2$\sigma$ (Figure \ref{fig:mike_galah_fig}). Based on MIST isochrones and single-star evolutionary models, we determine the red giant's mass, radius, and luminosity to be $0.99^{+0.02}_{-0.03}\,M_{\odot}$, $54.8 \pm 0.7,R_{\odot}$, and $518^{+13}_{-15}\,L_{\odot}$, respectively. We find that the observed ellipsoidal variability amplitude, when combined with the SED constraints, implies an orbital inclination of $(82^{+6}_{-7})^{\circ}$ (Figure \ref{fig:a2_fig}).
    
    \item We perform follow-up with the FEROS and MIKE high-resolution spectrographs (Figure \ref{fig:mike_galah_fig}) and measure radial velocities over more than one orbital cycle (Figure \ref{fig:rv_fig}). By fitting a Keplerian two-body orbit, we determine the orbital period of IGR J16194-2810 to be $192.73 \pm 0.01$ days. We derive a companion mass of $1.23^{+0.05}_{-0.03} ~ M_{\odot}$ (Figure \ref{fig:corner_plot} and Table \ref{tab:derived_params}), dynamically confirming the presence of a neutron star in the symbiotic X-ray binary.
    
    \item Informed by the Galactic orbit of IGR J16194-2810 (Figure \ref{fig:galactic_orbit}), we perform simulations to determine the distributions of allowed initial He star masses, orbital periods, and natal kicks that can reproduce the orbital parameters observed today (Figure \ref{fig:sn_params}). We also simulate the chance that the binary remains bound and at least as wide as observed following the supernova of the NS progenitor as a function of mass loss and natal kick (Figures \ref{fig:kick_sim} and \ref{fig:kick_constraints}), finding a non-zero probability of reproducing the current orbit for most reasonable initial configurations. We simulate the future evolution of the binary using MESA (Figure \ref{fig:donor_evolution}), concluding that the system will end up as a $\sim 0.4\,M_{\odot}$ helium WD orbiting the NS in a $\sim 900$ day orbit.
    
    \item Comparing IGR J16194-2810 (and other SyXBs) to the population of wide NS + MS binaries discovered in \textit{Gaia} DR3, we conclude that SyXBs with $P_{\rm orb} \lesssim 1000$\,d likely have a space density that is lower by a factor of $\gtrsim 100$ than wide NS + MS binaries. However, their X-ray bright lifetime is $\sim 1000$ times shorter than the lifetime of typical NS+MS binaries, implying a comparable or higher birth rate. As a result, the population of SyXBs provides a window on the future evolution of these unexpectedly wide NS binaries.
\end{itemize}

\subsection{Comparison with \citet{hinkle_2024}}
\label{sec:hinkle}

In a contemporaneous analysis, \citet{hinkle_2024} utilize optical and near-infrared spectroscopy to derive a spectroscopic orbit for IGR J16194-2810. They also analyze the ellipsoidal variability of the red giant and infer its stellar parameters and abundances. While they do not jointly fit all available data as we do in Section~\ref{sec:joint}, they derive orbital elements and stellar parameters in good agreement with our work. We compare our constraints against theirs in Table \ref{tab:hinkle_comparison}. 

\citet{hinkle_2024} do not attempt to measure the NS mass. We constrain it to $M_{\text{NS}} = 1.23^{+0.05}_{-0.03}\,M_{\odot}$. They assume the red giant exactly fills its Roche lobe when inferring its mass, but our analysis finds that it is only $\approx 82\%$ Roche lobe filling and that mass transfer is primarily via winds. The long spin period inferred from the TESS light curve \citep[$P\approx 4.06$ hours;][]{luna_2023} supports this interpretation, since mass transfer by Roche lobe overflow is expected to rapidly spin up the NS.

In contrast to our SED fitting and interpolation on MIST stellar models, \citet{hinkle_2024} measure a metallicity directly from near-IR spectra. Their result of $\rm [Fe/H] = -0.14 \pm 0.12$ is in excellent agreement with our value of $\rm [Fe/H] = -0.12^{+0.02}_{-0.01}$.  

\citet{hinkle_2024} also model the evolution of the system into a NS-WD binary. While they share our conclusions about the ultimate fate of IGR J16194-2810, they assume that the NS formed from a WD via accretion-induced collapse. We are agnostic about the formation process and show that system could have survived a NS natal kick of $\lesssim 50$ km s$^{-1}$. Furthermore, we provide additional context elucidating the connection between this SyXB and the population of wide NS-luminous star binaries discovered with \textit{Gaia} astrometry.

\begin{deluxetable*}{ccc}
\tablecaption{Comparison of orbital and stellar parameters derived in this work and \citet{hinkle_2024}. \label{tab:hinkle_comparison}}
\tablehead{\colhead{Parameter} & \colhead{Median Constraint (This Work)} & \colhead{ Constraint \citep{hinkle_2024}} \\
\colhead{(1)} & \colhead{(2)} & \colhead{(3)}}
\startdata
$P$ & $192.73 \pm 0.01$ days & $192.47 \pm 0.13$ days \\
$i$ & $(82^{+6}_{-7})^{\circ}$ & $(55 - 70)^{\circ}$\\
$T_0$ & $2460038.06 \pm 0.06$ & $2459026.33 \pm 0.33$\\
$\gamma$ & $-4.21 \pm 0.05$ km s$^{-1}$ & $-4.36 \pm 0.13$ km s$^{-1}$ \\
$K$ & $26.3 \pm 0.1$ km s$^{-1}$ & $25.58 \pm 0.16$ km s$^{-1}$ \\
$f(m)$ & $0.365 \pm 0.003\,M_{\odot}$ & $0.3337 \pm 0.0062\,M_{\odot}$\\
$M_{\text{NS}}$ & $1.23^{+0.05}_{-0.03}\,M_{\odot}$ & $1.2-1.5\,M_{\odot}$ \\ 
$M_*$ & $0.99^{+0.02}_{-0.03}\,M_{\odot}$ & $0.91 \pm 0.07 \,M_{\odot}$ \\ 
$R_*$ & $54.8 \pm 0.7\,R_{\odot}$ & $58\pm7\,R_{\odot}$ \\
$L_*$ & $518^{+13}_{-15}\,L_{\odot}$ & $573^{+72}_{-79}\,L_{\odot}$ \\
$T_{\text{eff}}$ & $3723^{+6}_{-7}$ K & $3700 \pm 100$ K \\
$\log \left(g/\text{cm s$^{-2}$}\right)$ & $0.958^{+0.004}_{-0.009}$ & $0.7-1.0$ \\
$\rm[Fe/H]$ & $-0.12^{+0.02}_{-0.01}$ & $-0.14 \pm 0.12$ \\
$v \sin i $ & $14.1 \pm 0.2$ km s$^{-1}$ & $18 \pm 2$ km s$^{-1}$ \\
\enddata
\end{deluxetable*}

\begin{acknowledgments}
We thank the referee for useful feedback. 
We are grateful to  Hans-Walter Rix and Vedant Chandra for coordinating FEROS observations to enable the data presented here to be collected, and to Tom Wagg for advice regarding orbit integration. 
This research was supported by NSF grant AST-2307232. This paper includes data gathered with the 6.5 meter Magellan Telescopes located at Las Campanas Observatory, Chile. This work has made use of data from the European Space Agency (ESA) mission
{\it Gaia} (\url{https://www.cosmos.esa.int/gaia}), processed by the {\it Gaia}
Data Processing and Analysis Consortium (DPAC,
\url{https://www.cosmos.esa.int/web/gaia/dpac/consortium}). Funding for the DPAC
has been provided by national institutions, in particular the institutions
participating in the {\it Gaia} Multilateral Agreement.


\end{acknowledgments}

%

\vspace{5mm}
\facilities{FEROS (La Silla), MIKE (Las Campanas)}


\software{astropy \citep{2013A&A...558A..33A,2018AJ....156..123A}, \texttt{emcee} \citep{emcee_2013}, \texttt{isochrones} \citep{isochrones_zenodo}, \texttt{gala} \citep{adrian_price_whelan_2024_10449846}}





\bibliography{bibliography}{}
\bibliographystyle{aasjournal}



\end{document}